\documentclass[aps,pra,twocolumn,superscriptaddress]{revtex4-2}

\usepackage{graphicx}
\graphicspath{ {./Figures/} }  
\usepackage{svg}  
\usepackage{xcolor}
\usepackage[normalem]{ulem}

\usepackage{bm}  
\usepackage{amsmath,amsfonts,amssymb}

\DeclareMathOperator{\arcsinh}{arcsinh}

\DeclareMathOperator{\tr}{tr}
\DeclareMathOperator{\Reel}{Re}

\begin{document}


\title{Exploring non-Euclidean photonics: Pseudosphere microlaser }

\author{H. Girin}
\email{hugo.girin@cnrs.fr}
\affiliation{Université Paris-Saclay, CNRS, Centre de Nanosciences et de Nanotechnologies, 91120, Palaiseau, France.}

\author{S. Bittner}
\affiliation{Université de Lorraine, CentraleSupélec, Laboratoire Matériaux, Optique, Photonique et Systèmes (LMOPS), 2 rue Edouard Belin, 57070 Metz, France.}

\author{X. Checoury}

\author{D. Decanini}
\affiliation{Université Paris-Saclay, CNRS, Centre de Nanosciences et de Nanotechnologies, 91120, Palaiseau, France.}

\author{B. Dietz}
\affiliation{Center for Theoretical Physics of Complex Systems,
Institute for Basic Science, Daejeon, 34126, Republic of Korea.}
\affiliation{Basic Science Program, Korea University of Science
and Technology (UST), Daejeon 34113, Republic of Korea}

\author{A. Grigis}
\affiliation{Laboratoire d’Analyse, Géométrie et Applications, CNRS UMR
7539, Université Sorbonne Paris Cité, Université Paris 13, Institut Galilée,
99 avenue Jean-Baptiste Clément, 93430 Villetaneuse, France}

\author{C. Lafargue}

\author{J. Zyss}

\affiliation{Laboratoire Lumière, Matière et Interfaces (LuMIn),
CNRS, ENS Paris-Saclay, Université Paris-Saclay, CentraleSupélec,
91190 Gif-sur-Yvette, France.}

\author{C. Xu}
\affiliation{Department of Physics, The Jack and Pearl Resnick Institute
for Advanced Technology, Bar-Ilan University, 5290002, Ramat-Gan, Israel.}

\author{P. Sebbah}
\affiliation{Department of Physics, The Jack and Pearl Resnick Institute
for Advanced Technology, Bar-Ilan University, 5290002, Ramat-Gan, Israel.}

\author{M. Lebental}
\email{melanie.lebental@cnrs.fr}
\affiliation{Université Paris-Saclay, CNRS, Centre de Nanosciences et de Nanotechnologies, 91120, Palaiseau, France.}
\affiliation{ENS Paris-Saclay, 91190  Gif-sur-Yvette, France.}

\begin{abstract}
	Classical and wave properties of microlasers with the shape of a truncated pseudosphere are
    investigated through experiments and numerical simulations. These pseudosphere microlasers are
    surface-like organic microlasers with constant negative curvature, which were fabricated with
    high optical quality by direct laser writing. It is shown that they behave, in many ways,
    similar to two-dimensional flat disks, regardless of their different Gaussian curvature.
    We derive the monodromy matrices for geodesics on the pseudosphere and demonstrate that
    the periodic geodesics are marginally stable. Actually, due to the rotational symmetry,
    the pseudosphere is an integrable system.
\end{abstract}

\maketitle

\section{Introduction}
\label{sec_1}

\begin{figure}
    \includegraphics[width = 1\linewidth]{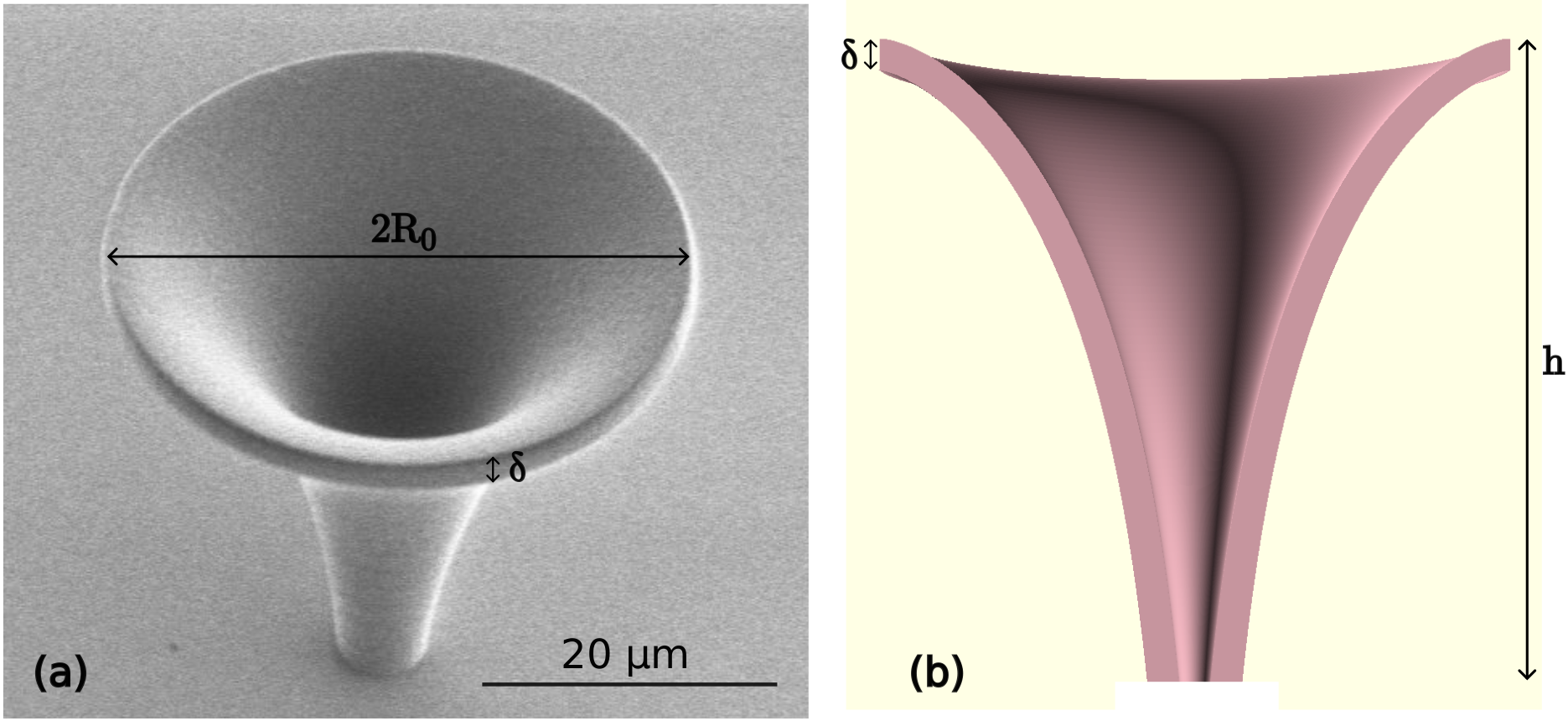}
    \caption{\label{fig:SEM_tractoid} (a) SEM image of a tractoid microlaser
    with radius $R_0 = 20$ µm and thickness $\delta = 2$ µm. (b) Schematic of a half tractoid microlaser with
    the same radius and thickness. The cavity with finite thickness $\delta$ is obtained by vertically
    shifting two tractoid surfaces with respect to each other. The height of the microlaser is around
    $h = 40$ µm.}
\end{figure}

This work on non-Euclidean photonics explores the ray and wave dynamics on the pseudosphere, a surface with
constant \emph{negative} curvature. Actually surfaces with \emph{positive} curvature, like the sphere, the
bottle \cite{bottle}, or the exterior of a toroid \cite{tore} have already given rise to an abundant
literature in photonics because of their relative ease of fabrication from glass by melting. On the
contrary, negatively curved surfaces - a field of research also called \emph{hyperbolic geometry} -
were intensively investigated by mathematicians \cite{Poincare1895}, theoretical physicists
\cite{voros, sieber, bogomolny, szepfalusy,quantumhall}, and artists \cite{escher},
while very few experiments were carried out  \cite{peschel-PRL,arie-polariton}
until the recent rise of three-dimensionnal (3D) printing at the microscale by Direct Laser Writing
(DLW) \cite{segev,Moebius}. This mature technology contributes by now to photonics, as illustrated by
the pseudospherical microlaser shown in Fig. \ref{fig:SEM_tractoid}.\\

In photonics, negatively curved surfaces provide two particular properties:
\begin{itemize}
\item On a curved surface light no longer propagates along straight lines, but along geodesics, which are the shortest paths between two points.
\item On a negatively curved surface each geodesic is unstable \cite{gutkin,instabilite}. That means that any pair of nearby trajectories will diverge exponentially from each other, which is the signature of chaotic dynamics.
\end{itemize}

In this paper we resort to two simplifications by choosing curved surfaces with the following properties:
\begin{itemize}
\item The curvature is certainly negative, but moreover constant, which allows to analytically calculate the stability of geodesics, as outlined in Sec. \ref{sec_4}, and gives access to the Poincaré half-plane (see Sec. \ref{subsec_2_2}), which is a very convenient parametrization for studying hyperbolic geometry.
\item We consider a truncated surface of revolution, which preserves rotational symmetry, implying that the
angular momentum is conserved, in addition to the energy. As propagation confined to the surface is a
two-dimensional (2D) dynamical system, there are as many constants of motion as degrees of freedom,
implying an integrable dynamics.
\end{itemize}
Among all the possible surfaces with these properties \cite{kuehnel}, we focus on the tractoid defined in
Sec. \ref{sec_2}. This surface has already been widely investigated, in particular in
Ref.~\cite{voros}. Our main contributions to this field are the stability ($\equiv$ monodromy) matrices
for propagation and reflection exhibited in Sec. \ref{sec_4}, and derived in App. \ref{appendix:stability}.
The demonstration is similar for classical massive particles and for light rays, as they basically follow
the same equations \cite{transmission}. We also show that each periodic geodesic is marginally stable, as
expected for a classically integrable dynamics.\\
On the wave-dynamical level (both matter waves or electromagnetic waves), the rotational invariance makes
the wave equation separable on the tractoid surface. Then the radial part of the Helmholtz equation
(for light) is mathematical equivalent to a Schrödinger equation with an effective
potential, which can be solved either analytically (Sec. \ref{sec:analytique-comparaison}), or
numerically (Sec. \ref{sec:num-dirichlet}), showing intriguing similarities to the flat circular
disk. Moreover experiments reported in Sec. \ref{sec_3} evidence lasing emission from Whispering
Gallery Modes (WGMs), like in most circular microresonators. Finally Sec. \ref{sec_6} concludes
the comparison between the flat disk and the tractoid with a full 3D numerical analysis taking
into account the actual thickness and polarization.

\section{Tractoid and Poincaré half-plane}\label{sec_2}
This first section introduces the tractoid surface and its representation in the Poincaré half-plane.

\begin{figure}
    \includegraphics[width=7.5cm]{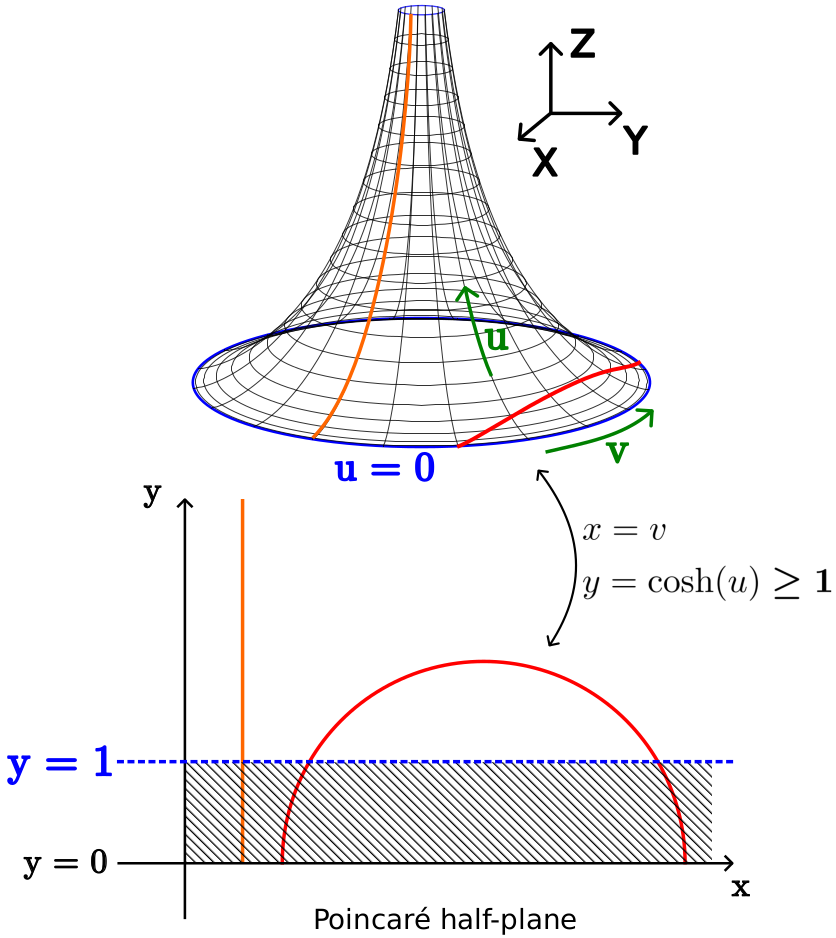}
    \caption{\label{fig:tractoid_PHP_v2} Transformation from the tractoid to the
    Poincaré half-plane. Two kinds of geodesics corresponding in the Poincaré half-plane either to a vertical line (orange)
    or to a half-circle whose center is on the $x$-axis (red) are sketched
    in both geometries. The blue circular lower boundary of the tractoid is mapped
    into the horizontal line $y=1$. Finally, the hatched area cannot be mapped on
    the tractoid.}
\end{figure}

\subsection{Parametrization of the tractoid}
\label{subsec_2_1}
A pseudosphere (or pseudospherical surface) is a
two-dimensional (2D) surface embedded in the 3D-Euclidean space with constant
negative Gaussian curvature. The simplest example of  a pseudosphere is
the tractoid, which is studied hereafter. A tractoid of radius $R_0$ is a surface having a Gaussian curvature
$-\frac{1}{R_0^{2}}$ at each point. For obvious similarities with the
sphere of radius $R_0$, whose Gaussian curvature is
$+\frac{1}{R_0^{2}}$, the tractoid is commonly referred to as a
pseudosphere in the literature. A typical parametrization of the tractoid
surface $\vec{r} = \vec{F}(u, v)$ is:
\begin{align}
    X &= F_X(u, v) = R_0 \frac{\cos(v)}{\cosh(u)} \label{eq:Fx}\\
    Y &= F_Y(u, v) = R_0 \frac{\sin(v)}{\cosh(u)} \label{eq:Fy}\\
    Z &= F_Z(u) = R_0[u-\tanh(u)] \label{eq:Fz}
\end{align}
where $R_0$ is the radius of the tractoid, $u \in [0, +\infty)$ is the
parameter along the generatrix and $v \in
[0, 2\pi)$ is the azimuthal angle (cf. Fig.
\ref{fig:tractoid_PHP_v2}). With this choice of local coordinates $(u, v)$, the
metric tensor has a form which is common for all Surfaces Of
Revolution (SORs)
\begin{equation}
    ds_{SOR}^{2} = E(u)du^{2}+G(u)dv^{2}
\end{equation}
where $E$ and $G$ are related to the radius of revolution
$R = \sqrt{X^{2} + Y^{2}}$ and to $Z$ as
\begin{align}
    G(u) &= R^{2}(u) \\
    E(u) &= \left(\frac{dR}{du} \right)^{2} + \left(\frac{dZ}{du} \right)^{2}
\end{align}
For the specific case of the tractoid, we find
\begin{align}
    G(u) &= \frac{R_0^{2}}{\cosh^{2}u} \label{eq:G}\\
    E(u) &= R_0^{2} \,\tanh^{2}u \label{eq:E}
\end{align}

Even though the tractoid is the simplest example of surface with constant
negative curvature embedded in $\mathbb{R}^{3}$, computations can become
tedious and it is generally more convenient to work in the Poincaré half-plane, which is depicted in Fig.
\ref{fig:tractoid_PHP_v2} and further described in the next paragraph.

\subsection{The Poincaré half-plane}
\label{subsec_2_2}
The Poincaré half-plane representation corresponds to the upper half-plane
$\mathbb{H} = \{(x,y) \in \mathbb{R}^{2} \mid y>0\}$ endowed with the Poincaré
metric
\begin{equation}
    ds_{HP}^{2} = R_0^{2}\,\frac{dx^{2} + dy^{2}}{y^{2}}
\end{equation}
The transformation
\begin{eqnarray}
\left([0\,;\,+\infty)\,,\,[0\,;\,2\pi)\right)&\longrightarrow &\mathbb{H}\\\nonumber
    (u, v) &\longmapsto & (x, y) = [v, \cosh(u)]
    \label{eq:transfo_PHP}
\end{eqnarray}
which maps the tractoid onto the subspace $\mathbb{A} = \{(x,y) \in
\mathbb{R}^{2} \mid y \geq 1\}$ of $\mathbb{H}$, preserves both the angles (conformal mapping) and
the distances (isometry). Therefore, both the geodesic equation and the wave
equation are conserved under this mapping \cite{PNASChenni}. In particular,
this means that distances can be computed  on the tractoid as well as
in the Poincaré half-plane. The length of a geodesic starting at $(x_1,y_1)$ and ending at $(x_2,y_2)$ equals \cite{voros}
\begin{equation}
L=2R_0\arcsinh \left(\frac{\sqrt{(x_1-x_2)^2+(y_1-y_2)^2}}{2\sqrt{y_1y_2}}\right)
\end{equation}
Note that the edge $u=0$ of the tractoid is mapped to the line $y=1$ (blue line
in Fig. \ref{fig:tractoid_PHP_v2}), and only the part of  the Poincar\'e plane above $y=1$ is accessed
by the transformation Eq.~(\ref{eq:transfo_PHP}).

The main advantage of working in the Poincar\'e half-plane is that the functional dependence of the
geodesics on the coordinates is simple. Indeed, there are only two kinds of geodesics.
These are either parts of semicircles, of which the centers are on the x-axis, or they are straight
lines $x=const.$ perpendicular to the x-axis, as shown in Fig.~\ref{fig:tractoid_PHP_v2}. Therefore
most of our calculations will be done in the Poincaré half-plane.

\section{Tractoid microlaser and experimental results}
\label{sec_3}
In this section we briefly explain the fabrication of the tractoid microlasers by Direct Laser Writing (DLW), the experimental setup, and exhibit results regarding the analysis of experimental laser spectra.

\subsection{Tractoid fabrication and experimental setup}
\label{subsec_3_1}

In the experiment, we designed a curved waveguide in the shape of the tractoid by shifting two tractoid
surfaces of radius $R_0=20$ µm with respect to each other. The distance between the two tractoid
surfaces, $\delta= 2\,\mu$m, defines the thickness of the tractoid microcavity [cf. Fig.
\ref{fig:SEM_tractoid}(b)]. It was cut at $u_{max} = 3$ which corresponds to a
height of about 40 $\mu$m.
The 3D design of the tractoid microlasers was created using the free OpenSCAD.
We would like to stress at this point that the tractoid microcavity differs from a mathematical
tractoid surface because of its finite size, which is accounted for in the computations by
considering truncated tractoids. Another difference is the finite thickness of this and other
curved microcavities. Yet, they can still be described by the scalar Helmholtz equation using
the concept of the effective index of refraction \cite{Moebius}, as outlined below.

The tractoid microlasers were then fabricated by DLW lithography using the
Photonic Professional GT+ system with
negative resist IP-G from the company Nanoscribe. The resist
was doped by 0.5 wt$\%$ pyrromethene 597 laser dye (by the company Exciton)
which is homogeneously distributed in the bulk host resist. A Scanning
Electron Microscope (SEM) image of a tractoid microlaser is
shown in Fig. \ref{fig:SEM_tractoid}(a).\\

\begin{figure}
    \includegraphics[width=9.0cm]{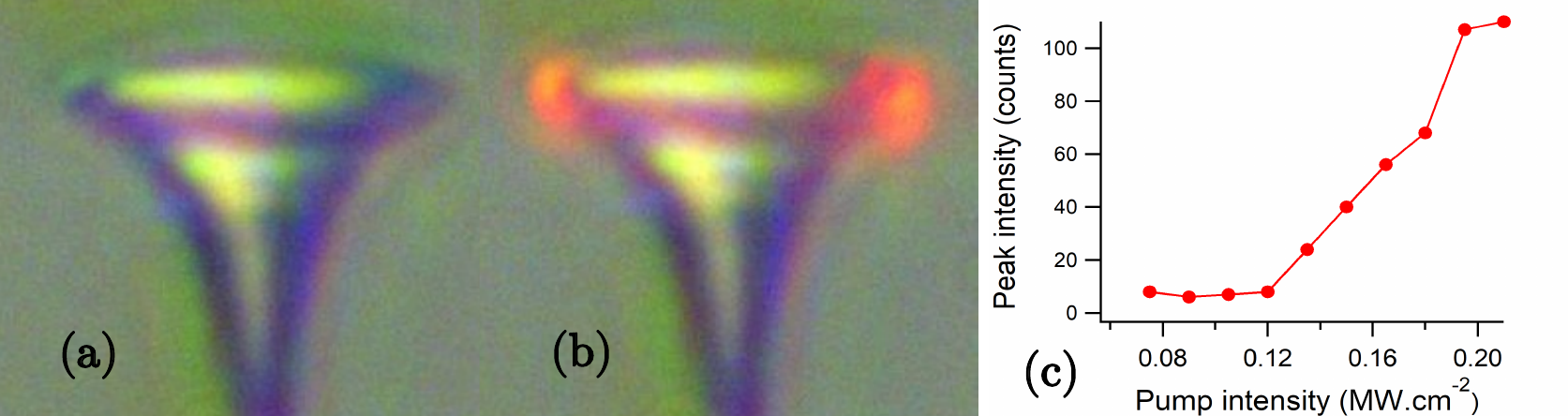}
    \caption{\label{fig:lasing_trumpet}Photographs of a tractoid microlaser
    illuminated with a low intensity white light without pumping (a), and
    with pumping (b).  The green pump
	light is removed by a notch filter, while the laser emission is orange. (c) Intensity of the peak at
    601.56 nm versus the pump power. The threshold is about 0.12 MW.cm$^{-2}$. Above  0.2 MW.cm$^{-2}$
    the curve has an irregular increase (not shown), which is  a signature of mode competition.}
    \label{fig:camera}
\end{figure}

The tractoid microlasers were pumped individually and with uniform
intensity by a beam perpendicular to the substrate from a frequency-doubled
Nd:YAG laser (532 nm, 500 ps, 10 Hz). Their emission was analyzed by a
spectrometer coupled to a CCD camera, providing an overall resolution of 0.03 nm.
The experiments were carried out at room pressure and temperature.
The microlaser light was mostly emitted parallel to the substrate plane.
The photograph in Fig. \ref{fig:camera}(b) shows a tractoid microcavity under pumping. The laser emission
mainly originates from the circular boundary at a near grazing angle and is observed from all directions,
which is consistent with the rotational symmetry of the microcavity. These observations
are reproducible from sample to sample, thusindicating that the lasing modes are mainly WGMs as confirmed
by the analysis of the laser emission spectra presented in the next section. A typical comblike laser
spectrum is shown in Fig \ref{fig:tractoid_exp_result}(a).\\
The threshold curve plotted in Fig. \ref{fig:camera}(c) evidences the laser effect. The
laser threshold is about 0.12 MW.cm$^{-2}$, which is one order of magnitude lower than for microlasers of
similar size and materials, evidencing the high quality factor of WGM.\\
Moreover the tractoid microlaser emission is polarized with the electric field oriented parallel to the
substrate, which corresponds to Transverse Electric polarization (TE) and will be discussed further in
Sec. \ref{sec_6}.

\subsection{Spectrum analysis}
\label{subsec_3_2}

\begin{figure}
    \includegraphics[width=7.0cm]{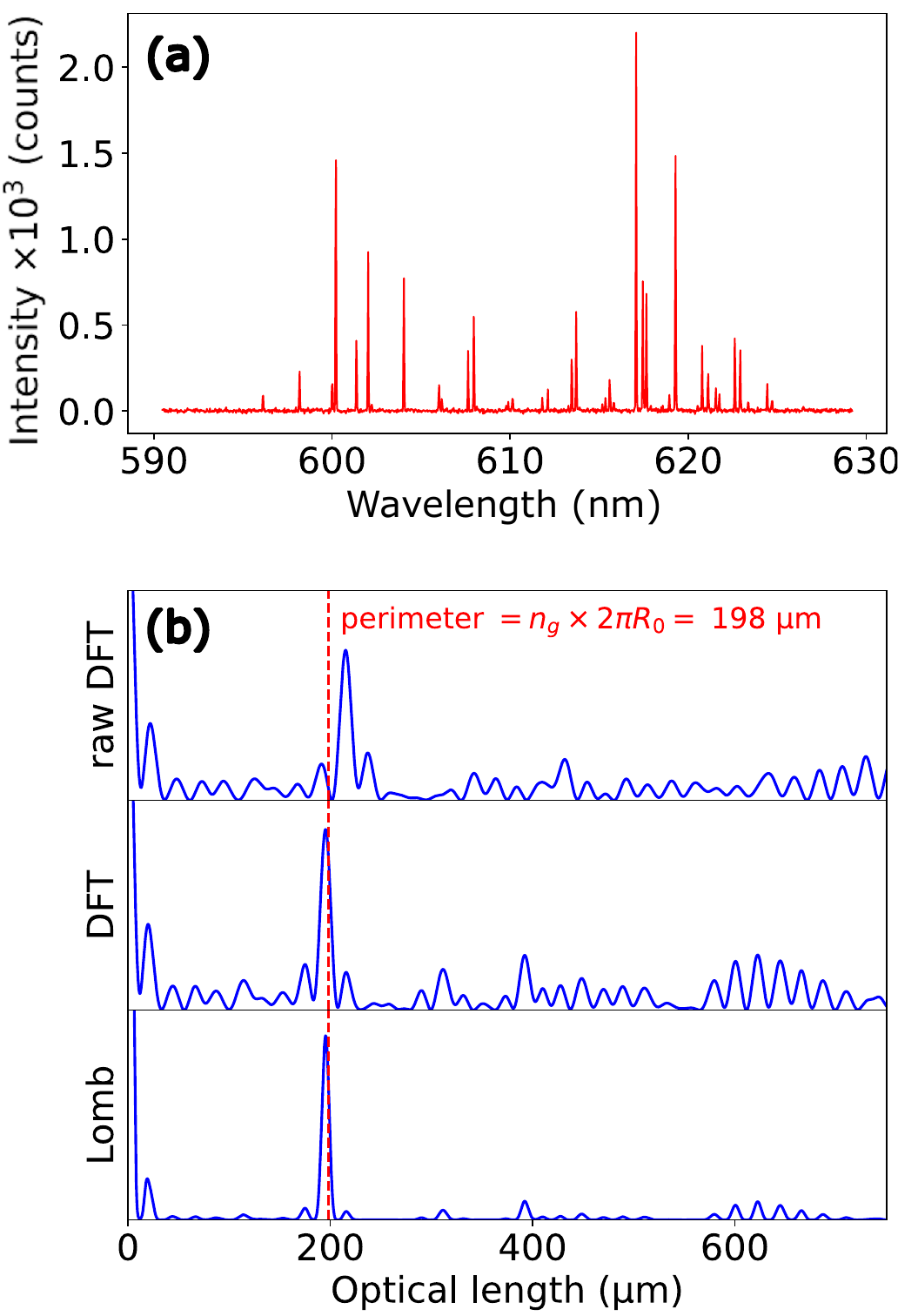}
    \caption{\label{fig:tractoid_exp_result} (a) Experimental laser spectrum of
    a tractoid microlaser. (b) Normalized Fourier
	transform of the spectrum in (a) from wavenumber to optical length using three different numerical
    methods: (top) Discrete Fourier Transform (DFT) algorithm which assumes that the sampling is regular,
    while the actual experimental data are not regularly sampled, (middle) same DFT algorithm, but taking
    into account the actual spacing between data points, and (bottom) Lomb transform. The perimeter of the
    tractoid microlaser is indicated by the red dashed line.}
\end{figure}

A typical experimental laser spectrum is presented in Fig. \ref{fig:tractoid_exp_result}(a). It is composed of several frequency combs. For each comb, the free spectral range or distance between two adjacent wavenumbers $\Delta k$ is inversely proportional to the geometric length $L$ of the corresponding
classical periodic orbit, as may be inferred from the semiclassical approximation of the spectral density \cite{berry}
\begin{equation}\label{eq:fsr}
    \Delta k = \frac{2 \pi}{n_g L}
\end{equation}
Here $n_g$ is the group refractive index, whose calibration protocol is detailed in App. V of Ref.
\cite{Moebius} and yields  $ n_g = 1.58 \pm 0.05$. The meaning of the bulk, effective, and group
indices involved in this study is explained in App. \ref{sec:indice}.

In Fig. \ref{fig:tractoid_exp_result}(b), the Fourier transform of the spectrum was calculated via three different methods, which are discussed in detail in App. \ref{sec:Lomb}. Our conclusion is that  the position of the peak is most accurately determined with the Lomb method and moreover the result is less noisy.

The Fourier transform is peaked at $ n_gL = (193 \pm 6) \,\mu$m.  This length is slightly smaller than the perimeter of the tractoid $P = n_g2 \pi R_0 = (198\,\pm 10)\,$ µm
as shown by the dashed red line in Fig. \ref{fig:tractoid_exp_result}(b).
It thus corresponds to WGMs propagating along the
circular boundary of the microlaser at $u=0$, which is consistent with the observation of emission spots located on the boundary of the tractoid [cf. Fig. \ref{fig:lasing_trumpet}(b)], as discussed in Sec. \ref{subsec_3_1}.

\section{Stability of the periodic geodesics}
\label{sec_4}
This section deals with the classification and stability of periodic geodesics on the tractoid. Since the experiments were carried out in the semiclassical regime $kR_0\simeq 200$,  geometrical optics approach is relevant. In this regime, the spectral density is well approximated in terms of a sum over periodic geodesics with a weighting depending
on their stability \cite{brack}, and thus experimentally accessible through the Fourier transform of the spectra. We demonstrate here that each periodic geodesic of the full tractoid $u\in[0,+\infty[$ is marginally stable, as expected from the rotational invariance. The case of the truncated tractoid is similar and analyzed in App. \ref{subappendix:tractoide-tronquee}

\subsection{Classification of the periodic geodesics}
\label{subsec_4_1}

To determine the stability of the periodic geodesics of the full tractoid with $u\in[0,+\infty[$, we work in the Poincaré half-plane, since there
is a one-to-one correspondence between geodesics on the actual tractoid and
in the Poincaré half-plane where their expression is simpler (cf. Sec.
\ref{subsec_2_2}).

There exist only two families of geodesics, the vertical lines $x =\,$constant
and  arcs of circles with center on the $x$-axis $y=0$. A trajectory on the tractoid cannot switch from one kind of geodesics to
the other one. Indeed, when moving
along a straight-line geodesic, the ray is reflected back at
 $y = 1$, while a ray evolving on a circular-arc type
geodesic circle does not impinge the line $y = 1$ perpendicularly. Therefore, arc of
circles are the only way to obtain periodic geodesics.

\begin{figure}
    \includegraphics[width=8.5cm]{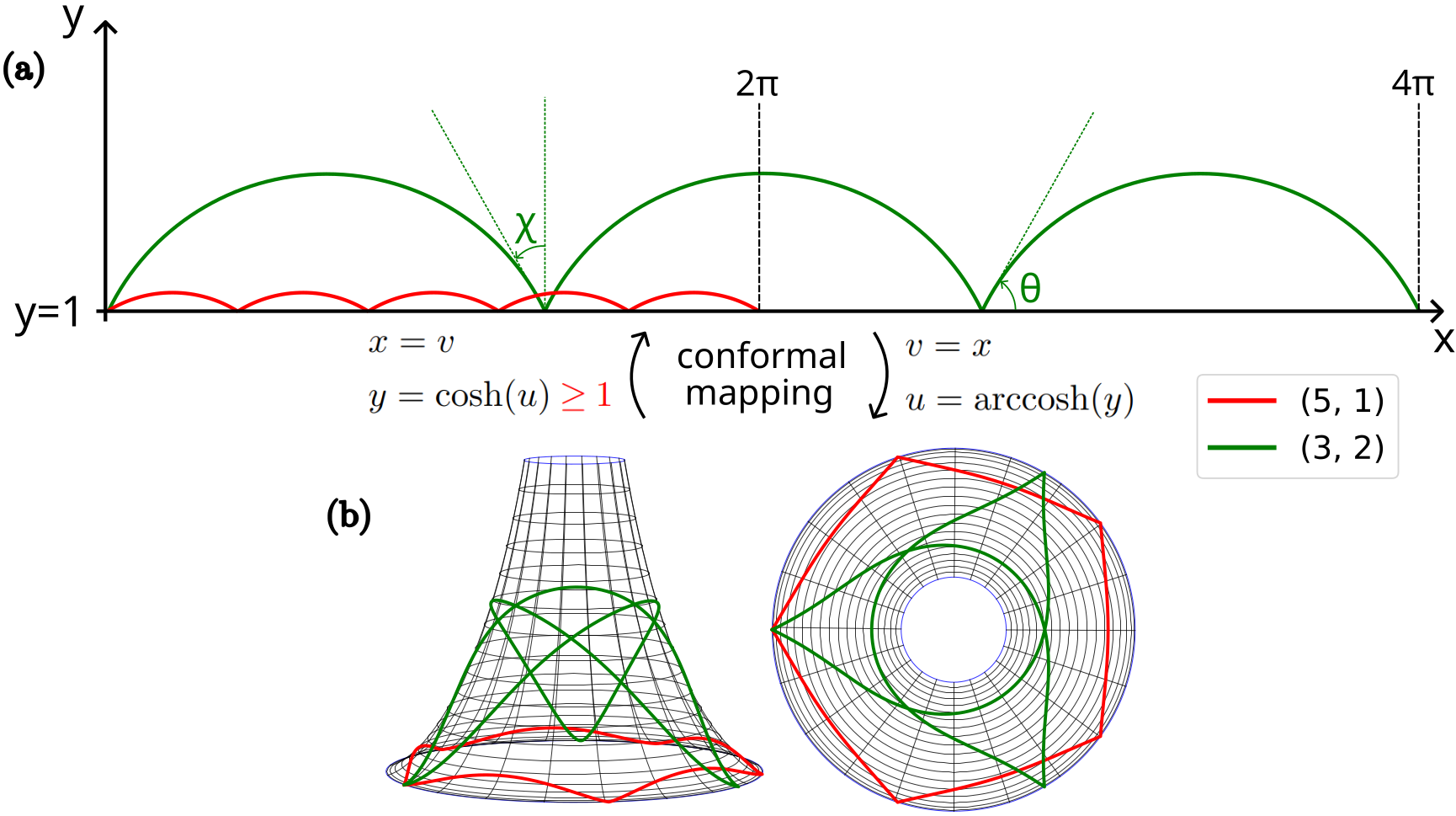}%
    \caption{\label{fig:schema_PG} Periodic geodesics with indices
    $(M, P) = (5, 1)$ (red) and $(M, P) = (3, 2)$ (green) (a) in the Poincaré half-
    plane and (b) on the tractoid.}
\end{figure}

Since the angles before and after reflection are
the same, the ray propagates along a periodic array of
identical  circular arcs. An example is depicted in Fig. \ref{fig:schema_PG}. Thus,
periodic trajectories are uniquely identified by a pair of
positive, relatively prime integers $(M,P)$ where $M$ is the
number of bounces on the boundary $y = 1$ and $P$ is the
winding number, that is, the  number of complete rotations
of $2\pi$ around the tractoid. Thus, the periodic geodesics on the tractoid have the same structure as periodic orbits of the flat disk.

The  length $L_{M, P}$ of the
$(M, P)$-periodic geodesic as well as $\theta_{M, P}$ the angle between
the line $y=1$ and the geodesic at the reflection point read (cf. Fig. \ref{fig:schema_PG}
for the notations):
\begin{align}
    L_{M, P} & = 2  M  R_0  \arcsinh \left( \frac{P}{M}\, \pi \right)
    \label{eq:length_PG} \\
    \tan \theta_{M, P} & =  \frac{P}{M} \,\pi
    \label{eq:angle_PG}
\end{align}
They will be used to determine the stability of the periodic geodesics in the next paragraph.

\subsection{Stability of the periodic geodesics}
\label{subsec_4_2}
The stability of the $(M, P)$-periodic geodesic depends on the eigenvalues
of its monodromy matrix $\mathcal{M}$, which describes the result of a small perturbation of the initial conditions on the trajectory (see App. \ref{appendix:stability} for the definition). Since the $(M, P)$-periodic geodesic
corresponds to $M$ repetitions of the same  circular arc, $\mathcal{M}$ is the
product $\mathcal{M} = (\mathcal{M}_0)^{M}$ where $\mathcal{M}_0=\mathcal{P}\,\mathcal{R}$ is the product of $\mathcal{P}$ the monodromy matrix for the propagation along a single circular arc, and $\mathcal{R}$ the monodromy matrix for reflection on the circular boundary of the tractoid.

\medskip
After computations based on Ref. \cite{sieber2} and detailed in App. \ref{appendix:stability}, the
monodromy matrix $\mathcal{P}(t)$ for the propagation along a geodesic over a distance $t$ is given by
\begin{equation}\label{eq:matrice-pi}
	\mathcal{P} (t) =
    \begin{pmatrix}
        \cosh \left( \frac{t}{R_0} \right) & R_0 \sinh \left( \frac{t}{R_0} \right) \\
        \frac{1}{R_0} \sinh \left( \frac{t}{R_0} \right) & \cosh \left( \frac{t}{R_0} \right)
    \end{pmatrix}
\end{equation}
This expression is relatively simple because the tractoid has a constant Gaussian curvature.
Then, the general formula for the monodromy matrix of reflection with incident angle $\chi$ of a ray propagating on a curved surface is
\begin{equation}\label{eq:matrice-R}
    \mathcal{R}(\chi) =
    \begin{pmatrix}
        -1 & 0 \\
        \frac{2 \epsilon\kappa_t}{\cos \chi} & -1
    \end{pmatrix}
\end{equation}
where $\epsilon=\pm1$, and  $\kappa_t$ is the modulus of the tangent curvature (also called geodesic curvature), which measures the curvature of the boundary in the plane which is tangent to the surface and is also the plane of incidence of the trajectory. General formulas for $\kappa_t$ and $\epsilon$ are given in App. \ref{subappendix:stability2}. For a reflection on the lower circular
boundary of the tractoid $u=0$, $\epsilon=+1$ and $\kappa_t= \frac{1}{R_0}$ .
Then using formulas (\ref{eq:length_PG}) and (\ref{eq:angle_PG}), one obtains (see App. \ref{subappendix:tractoide-complete}):
\begin{equation}
	\tr \left[ \mathcal{M}_0(M, P) \right] = \tr\left[\mathcal{P}(M, P) \mathcal{R}(M, P)\right] = 2
\end{equation}
As $\det \mathcal{M}_0=1$, the two eigenvalues of $\mathcal{M}_0$ are both equal to 1.
Therefore, like for the flat
disk, every periodic geodesic is  \textit{marginally stable}.
This might come as a surprise because the propagation itself is unstable, (ie. $\tr ( \mathcal{P})  > 2$. Yet, the focusing effect from the reflection at the boundary cancels the divergence during propagation. Of course, it is also a consequence of the rotational symmetry of the tractoid.

\begin{figure}
    \includegraphics[width=8.5cm]{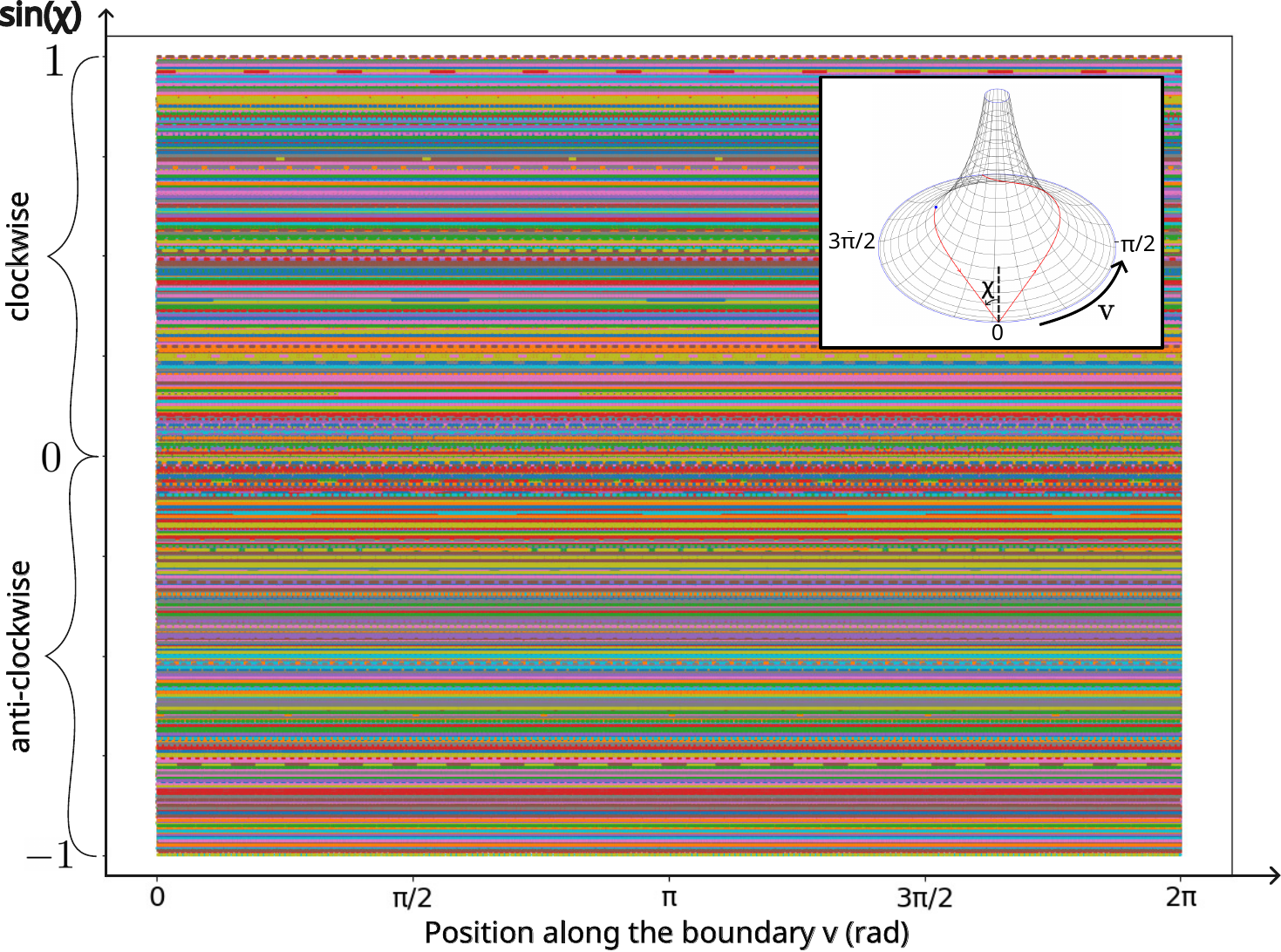}
    \caption{\label{fig:PSOS} Poincaré Surface of Section for the truncated
    tractoid ($u_{max} = 2$). Each color corresponds to a specific initial condition ($v$, $\chi$).  Inset:
    Definition of the incident angle $\chi$ and the curvilinear coordinate $v$ for the position on the boundary.}
\end{figure}

Finally, using a ray tracing
algorithm, we numerically compute  the Poincaré Surface of
Section (PSOS) of the tractoid. As shown by Fig. \ref{fig:PSOS}, every trajectory covers a horizontal line with $\sin\chi=$ constant, which is a manifestation of angular momentum conservation. The phase space structure is identical to that of a flat disk. There are neither stable islands nor a chaotic sea which is
typical for integrable billiard systems.

\medskip
To conclude this section,  every periodic geodesic
is marginally stable, as expected for a surface of revolution. It should be noted that the truncated tractoid ($u \in  [0, u_{max}]$)
is also a surface of revolution. While other periodic geodesics exist in this geometry, their stability is also neutral, as outlined in App. \ref{subappendix:tractoide-tronquee}.

\section{Solutions of the wave
equation with Dirichlet boundary conditions}
\label{sec_5}

The electromagnetic field obeys Maxwell's equations with a refractive index $n(\vec r)$ being  1.51 inside the tractoid \cite{indice} and 1 outside. The numerical simulations of Sec. \ref{sec_6} makes no additional assumption \cite{PML}. In this section,  we simplify the physical system to perform analytical calculations and provide a deeper understanding.

Following the "effective index approximation" for curved layers proposed in Ref. \cite{Moebius}, we assume that the light is well-confined
within the layer and propagates with an effective index $n_{eff}$, which
accounts for the varying refractive indices in the direction perpendicular to
the surface.  Furthermore Ref. \cite{peschel} demonstrated that the tractoid surface should support two
distinct polarization classes of modes, Transverse Electric modes or TE-modes (resp. Transverse Magnetic
modes or TM-modes) for which the electric (resp. magnetic) is parallel to the tractoid surface. Actually
all the necessary assumptions of Ref. \cite{peschel} are valid for the tractoid, namely it is a surface of
revolution, its main curvature $H$ is null, and the lasing wavelength is much smaller than the scale of the
curvature. Experimental results also indicate that the lasing modes of the tractoid microlaser are TE-modes
(with the electric field in the plane $Z=0$). Hence we can resort to the 2D scalar Helmholtz equation
\begin{equation}\label{eq:vec_helm2D}
    (\Delta_s + n_{eff}^2k^{2}) \psi = 0
\end{equation}
where $\Delta_s$ is the  Laplace operator restricted to the tractoid surface, and the wave function $\psi$ corresponds to the component of the magnetic (resp. electric) field normal to the surface for TE-modes (resp. TM-modes).

Due to rotational symmetry, the wave equation (\ref{eq:vec_helm2D}) is separable, and can be solved analytically. To simplify the analytical treatment, and later on facilitate the comparison with the flat disk, we impose Dirichlet boundary conditions, ie. $\psi(u=0,v)=0$, $\forall v\in[0,2\pi[$. In this case, $n_{eff}$ acts merely as a scaling factor, so we set $n_{eff}=1$.\\

In Sec. \ref{sec5_sub1}, we separate variables in order to write Eq. (\ref{eq:vec_helm2D}) as a Schrödinger-like equation along the radial coordinate. The localization of the wave at the boundary $u=0$ can then be interpreted through an effective potential. Then in Sec. \ref{sec:analytique-comparaison}, we derive the analytical solutions of Eq. (\ref{eq:vec_helm2D}), which are compared to the analytical solutions of the flat disk in Sec. \ref{sec:comparaison-disque}. As the analytical solutions of the tractoid are not easy to handle, we propose an efficient numerical method in Sec. \ref{sec:num-dirichlet} to provide a large number of eigenvalues. Finally we show that their distribution is consistent with an integrable system.

\subsection{Schrödinger-like equation}
\label{sec5_sub1}

\begin{figure}
    \includegraphics[width=6cm]{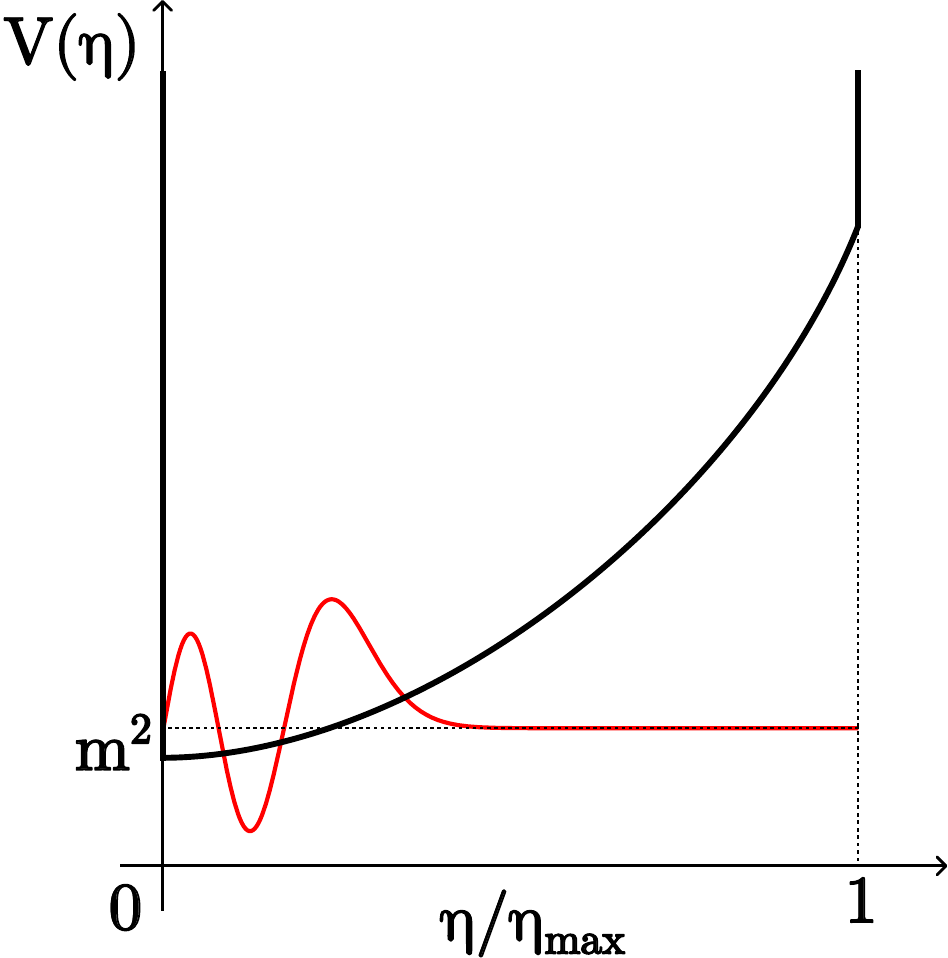}
	\caption{\label{fig:potential} Effective potential well, $V_{eff}^{(m)}(\eta) = m^{2} e^{2 \eta}$ for
    $m=60$ emerging from the Schrödinger-like equation (\ref{eq:schrodinger_eq}) for the truncated tractoid
    (black solid line). $\eta=0$ corresponds to the boundary $u=0$. The red line is the real-valued wave
    function corresponding to $m=60$ and $p=3$. With a $\mu^2$ offset, it oscillates around the horizontal
    dotted line which is positioned at the eigen-energy $\mu^2$.}
\end{figure}

The Laplacian operator is given by
\begin{equation}
\Delta_s\psi=\frac{1}{\sqrt g}\sum_{i,j=1}^2\frac{\partial}{\partial q_i}\left(
\sqrt{g}\,g^{ij}\frac{\partial \psi}{\partial q_j}\right)
\end{equation}
with $g_{ij}$ the metric tensor, $g^{ij}$ its inverse, and $g=\det (g_{ij})$. Hereafter, the derivation is
presented in the
Poincaré half-plane for simplicity. The equivalent derivation in the
actual tractoid is described
in App. \ref{appendix:wave_eq_SOR}. The metric
tensor of the Poincaré half-plane is
$g_{ij} = R_0^{2}\, y^{-2} \,\delta_{ij}$. Then, the Laplace operator
in the Poincaré half-plane is
\begin{equation}
\Delta_s = y^{2} \Delta_{xy}=\frac{y^2}{R_0^2}( \partial_x^{2} + \partial_y^{2})
\end{equation}
with $\Delta_{xy}=\partial_x^{2} + \partial_y^{2}$ the Laplace operator in the Euclidean plane. Hence, the wave equation becomes
\begin{equation}
    \left[ y^{2}\Delta_{xy} + (k R_0)^{2} \right]\psi = 0
    \label{eq:wave_eq_2D}
\end{equation}
This equation can be solved by variable separation using the ansatz $\psi(x,y) = \zeta(y) e^{imx}$ with $m \in \mathbb{Z}$. By inserting the expression of $\psi$ in the wave equation
(\ref{eq:wave_eq_2D}), we obtain the equation satisfied by $\zeta$
\begin{equation}
    -y^{2} \frac{d^{2}\zeta}{dy^{2}} + y^{2}m^{2}\zeta(y) = (k R_0)^{2} \zeta(y)
    \label{eq:wave_eq_1D}
\end{equation}
The $y^{2}$ term in front of the second derivative is removed with
the change of variable $y = e^{\eta}$. Then, the first-order derivative is
removed by introducing the function $\xi(\eta)$  defined by
\begin{equation}
    \zeta(\eta) = e^{\frac{\eta}{2}} \xi(\eta)
\end{equation}
Finally, we obtain a Schrödinger-like equation satisfied by $\xi$
\begin{equation}
    -\frac{d^{2}\xi}{d \eta^{2}} + m^{2} e^{2 \eta} \xi(\eta) =
    \mu^{2} \xi(\eta)
    \label{eq:schrodinger_eq}
\end{equation}
where $\mu^{2} = (kR_0)^{2} - \frac{1}{4}$. This equation corresponds to a
Schrödinger equation for a particle trapped in the potential well $V_{eff}^{(m)}(\eta) = m^{2} e^{2 \eta}$.
It is plotted for $m=60$ in Fig. \ref{fig:potential} together with the occupation probability density
$\psi(u,v)^2$ with $(m,p)=(60,3)$, which is similar to that for the corresponding flat disk near the
bottom of the effective potential, at the boundary $u=0$, ie. $\eta=0$.\\
Furthermore Eq. (\ref{eq:schrodinger_eq})  can be transformed into a Bessel equation with imaginary index,
as shown in the next paragraph.

\subsection{Analytic solutions}\label{sec:analytique-comparaison}

In this section, we exhibit analytical solutions of the 2D
wave equation for the tractoid \cite{szepfalusy} and compare them with the flat disk.

\begin{figure}
    \includegraphics[width=6cm]{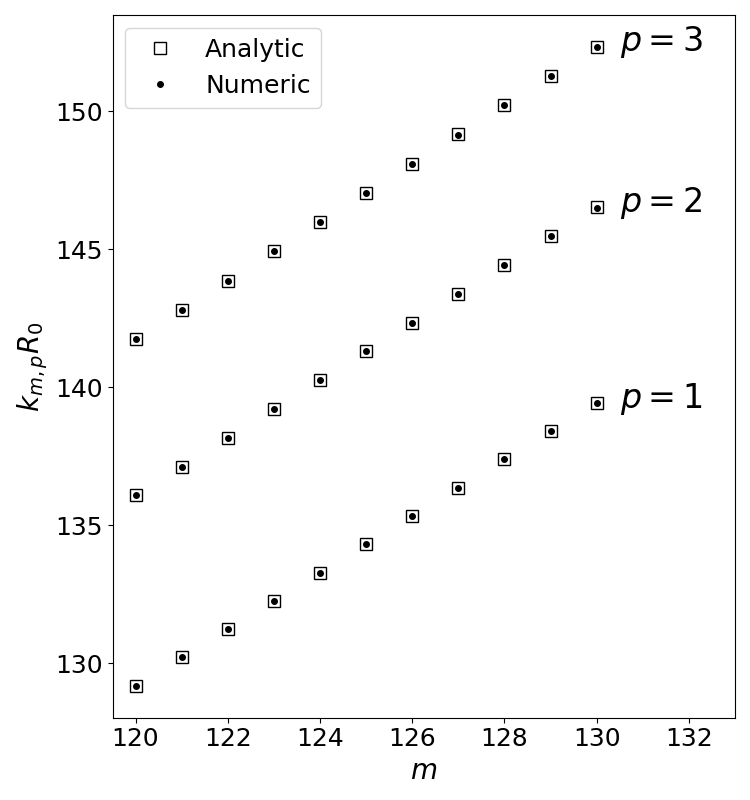}%
    \caption{\label{fig:mathematica} Comparison between the analytical
    spectrum of the tractoid computed with Mathematica (open squares) and the numerical
    spectrum computed using the matrix approach developed in
	Sec. \ref{sec:num-dirichlet} (dots). }
\end{figure}

We start from the wave equation (\ref{eq:wave_eq_1D}) obtained in the Poincaré half-plane after separation of the variables, and perform the variable change $w = m y$.
Then the function change $\zeta(w) = \sqrt{w} \,\gamma(w)$ leads to
\begin{equation}
	\frac{d^{2} \gamma (\omega)}{dw^{2}} + \frac{1}{w} \frac{d \gamma(\omega)}{dw}
	- \left[1 + \frac{\frac{1}{4}-(kR_0)^{2}}{\omega^2}\right]\gamma(w) = 0
\end{equation}
which is actually the equation for a modified Bessel function with index $\tilde\mu$, where
$\tilde\mu^{2} = 1/4-(kR_0)^{2}$ and $\tilde\mu=i\sqrt{(kR_0)^{2}-1/4}\equiv i\mu$, which is
purely imaginary for $kR_0 > 1/4$. This equation has two linearly independent solutions which
are called the \textit{modified Bessel function of the first and the second kind}
$K_{i \mu}$ and $I_{i \mu}$. They are related to the Bessel functions of the first and
the second kind by
\begin{align}
    I_{i \mu}(w) & = e^{ \mu \frac{\pi}{2}} J_{i \mu}(iw) \\
    K_{i \mu}(w) & = \frac{i \pi}{2} e^{- \mu \frac{\pi}{2}} \left[ J_{i \mu}(iw) + i Y_{i \mu}(iw) \right]
\end{align}
A general solution of the wave equation can thus be written as a linear combination of $I_{i \mu}$ and
$K_{i \mu}$. However, $I_{i \mu}(w)$ diverges for $w \to \infty$, while $K_{i \mu}(w)$
is finite. Therefore, if we consider the whole tractoid with $u
\to \infty$ (ie. $w \to \infty)$, only the $K_{i\mu}$
solution should remain:
\begin{equation}
    \psi(u, v) = C e^{i m v} \sqrt{m \cosh u}\, K_{i \sqrt{(k R_0)^{2} - 1/4}}(m \cosh u)
\end{equation}
with $C \in \mathbb{C}$ a constant. For simplicity, we consider  the Dirichlet boundary condition at $u=0$, that is
\begin{equation}
    K_{i \sqrt{(k R_0)^{2} - 1/4}}(m) = 0
    \label{eq:secular_eq}
\end{equation}
For a given integer $m$, the real $k$ values satisfying Eq. (\ref{eq:secular_eq}) are denoted as $k_{m,p}$ with $p=1,2,\ldots$ the radial order of the root.
Finding these roots $k_{m,p}$ is still an open problem
in spite of the broad bibliography. Recently, it
became possible to obtain an analytical solution of this equation for large
values of $k$ \cite{bessel}. Nevertheless, the matrix method developed in Sec.
\ref{sec:num-dirichlet} provides an efficient way to approximate these roots. Indeed,
the eigenvalues computed in Sec. \ref{sec:num-dirichlet} are compared to the roots of Eq.
(\ref{eq:secular_eq}) around $m=130$ computed by the \texttt{FindRoot} command on Mathematica
and plotted in Fig. \ref{fig:mathematica}. The agreement between both spectra is excellent,
with a relative difference smaller than $4 \times 10^{-6}$ for the first
branch $p = 1$ and around $2 \times 10^{-5}$ for the third branch $p=3$.\\

\subsection{Comparison with the flat disk}\label{sec:comparaison-disque}

\begin{figure}
    \includegraphics[width=7cm]{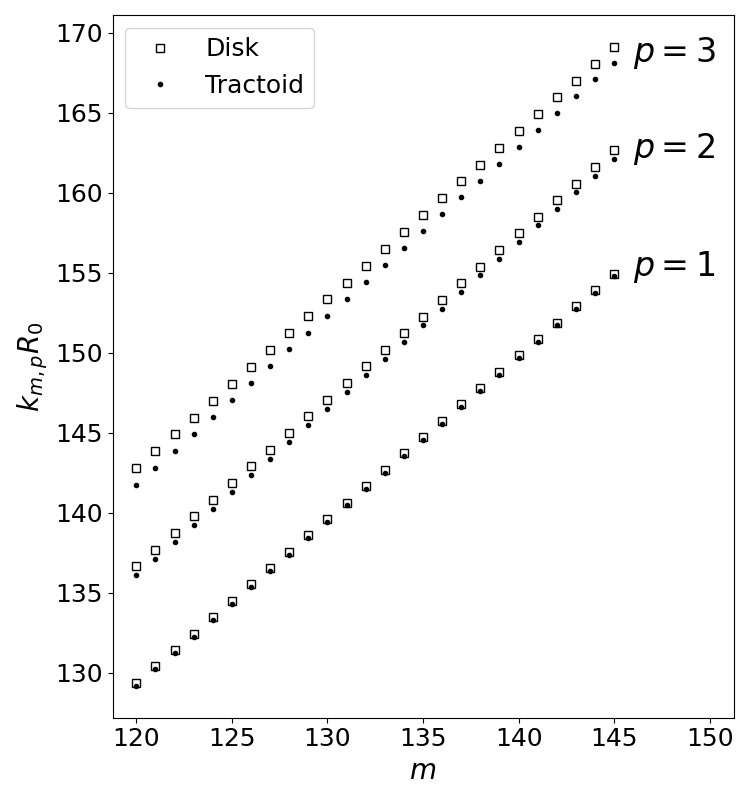}
    \caption{\label{fig:compa_disk_tractoid} Comparison between the spectrum of the
    flat disk (squares) and the spectrum of the tractoid (dots). The radial order $p$ is equal to
    1 (bottom), 2 (middle) and 3 (top). Only modes in the range $m=$ 120-145 are shown. The disk spectrum is calculated analytically and
    the tractoid spectrum is obtained by the method described in section \ref{sec:num-dirichlet}.}
\end{figure}

In the previous paragraph, we showed that the wave equation of the tractoid can be solved analytically. Its spectrum will be now compared with the spectrum of the flat disk.

Solutions of the radial wave equation for a disk of radius $R_0$ with Dirichlet boundary conditions are the Bessel function of the first kind, that is,
$\psi(r,\varphi)=J_m(kr)\phi(m\varphi)$, with $\phi(x)=\cos x$ or $\phi(x)=\sin x$ in polar coordinates $(r,\varphi)$ with $m \in\mathbb{N}$. The spectrum is real positive and
is given by the boundary conditions
\begin{equation}
    J_m(k R_0) = 0
    \label{eq:secular1}
\end{equation}
A major difference with Eq. (\ref{eq:secular_eq}) for the tractoid is that there the wavenumber $k$
appears in the index $i\mu$ of the modified Bessel function and not in the argument, as in
Eq.~(\ref{eq:secular1}).
The spectra of the disk and the tractoid are plotted in  Fig.~\ref{fig:compa_disk_tractoid}.
When $m$ is sufficiently large, and $p$ small, modes of the disk and the tractoid are very close to
each other. For $p = 1$ and $m\simeq 100$, the relative difference between modes of the disk and the
tractoid is around 0.2 \%. When $p$ gets larger, the difference also increases. Actually for a large
azimuthal number $m$ and a small radial number $p$, the wave is mainly located along the circular
boundary of the tractoid [cf. Fig. \ref{fig:analytic_trumpet}(a)] where the tractoid surface looks
similar to a flat disk.

\subsection{Numerical procedure to solve the wave equation}
\label{sec:num-dirichlet}

Since the roots of Eq. (\ref{eq:secular_eq}) are not easy to compute, we employed a convenient numerical
finite-element method to solve the Schrödinger-like Eq. (\ref{eq:schrodinger_eq}) for the effective
potential of Fig. \ref{fig:potential} corresponding to the tractoid, which is explained in App.
\ref{appendix:der_wave_eq}.

For comparison with previous paragraphs, we consider Dirichlet boundary conditions at $\eta=0$
(corresponding to $u=0$). As discussed before, the wave functions are expected to be localized
in the potential well at $\eta=0$, and  to be insensitive to the boundary condition at $\eta= \eta_{max}$
(corresponding to a tractoid truncated at $u=u_{max}$). Therefore to simplify, we also apply Dirichlet
boundary condition at $\eta=\eta_{max}$. To summarize, we have to solve Eq. (\ref{eq:schrodinger_eq})
with the Dirichlet boundary conditions    $\xi(\eta=0) = \xi(\eta_{max}) = 0$, where
$e^{\eta_{max}} = \cosh(u_{max})$.

After a discretization protocol explained in App. \ref{appendix:der_wave_eq}, we obtain a system of linear
equations that can be written with matrix notations as
\begin{equation}
    (\mathcal{T} + \mathcal{V}) \,\vec{\xi} = \mathcal{H} \vec{\xi} = \mu^{2} \vec{\xi}
    \label{eq:eigen}
\end{equation}
where the notation $\mathcal{H} = \mathcal{T} + \mathcal{V}$ has been used. $\mathcal{T}$ is a tridiagonal matrix
which accounts for the kinetic energy term described by expression
(\ref{eq:discret}) and the effective potential is represented by the
diagonal matrix $\mathcal{V}$. The operator $\mathcal{H}$ is then equivalent to the Hamiltonian
operator of a quantum mechanical system.

\begin{figure*}
    \includegraphics[width = 1\linewidth]{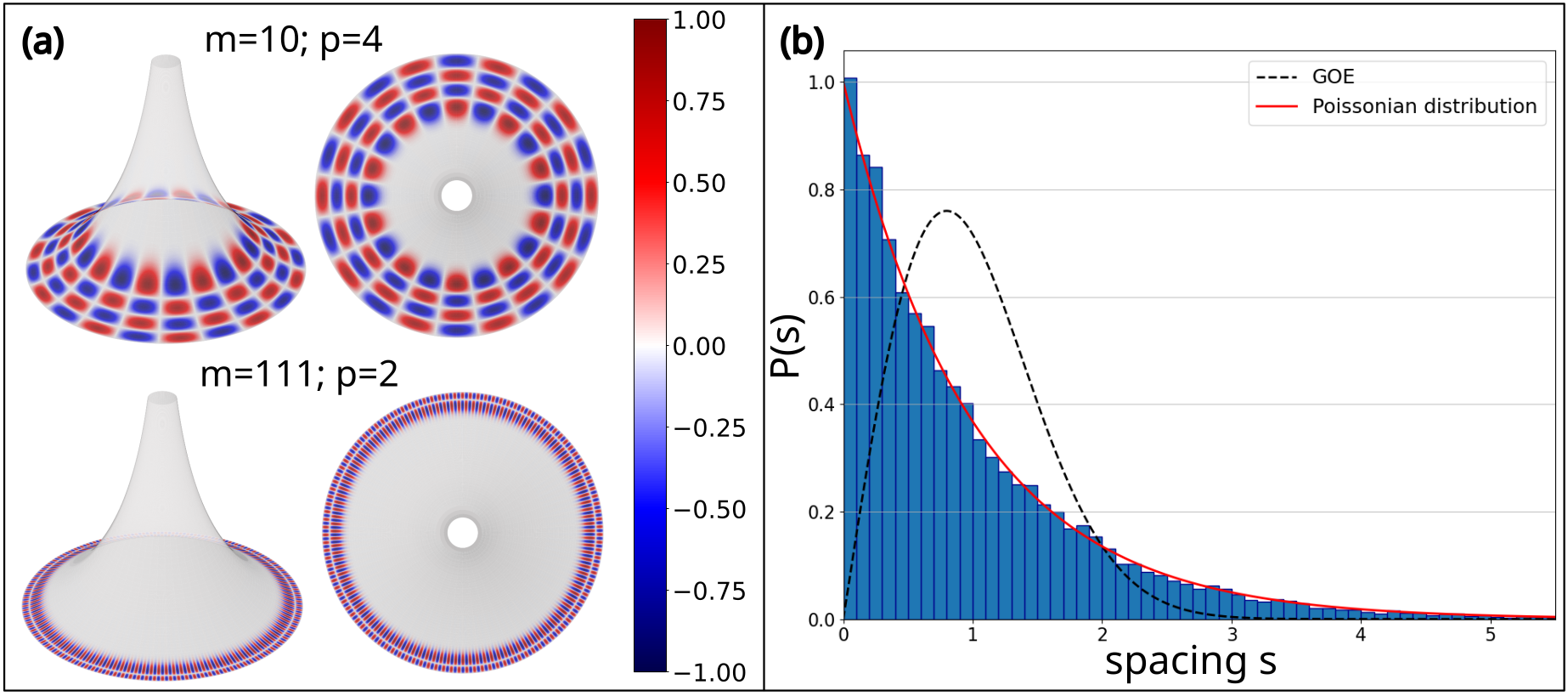}
	\caption{\label{fig:analytic_trumpet} (a) Different views of the $(10, 4)$-mode (top)
	and the $(111, 2)$-mode (bottom) calculated by the matrix approach of Sec. \ref{sec:num-dirichlet}. The wave functions are real-valued.
	(b)  Nearest-Neighbour Spacing Distribution calculated based on the
	4000 modes found between $kR_0 = 75$ and $k R_0 = 150$. The Wigner surmise for random matrices from the Gaussian
    Orthogonal Ensemble (GOE) is plotted with a dashed black line, and the Poissonian distribution with a solid red line.}
\end{figure*}

The problem of
finding the wave functions and the corresponding wavenumbers $k R_0$
has been turned into finding the
eigenvectors and the eigenvalues of the tridiagonal matrix $\mathcal{H}$.
It can be done by using the Relatively Robust Representations method implemented
on Python in the \texttt{scipy} library. Similarly
to the disk, modes are denominated by two integers, $m$ and $p$. For a given azimuthal number
$m \in \mathbb{Z}$, there are several solutions indexed by the radial number $p \in \mathbb{N}^{*}$. Some examples of  wave functions are shown in Fig. \ref{fig:analytic_trumpet}(a). They ressemble WGMs in a flat disk.

Using this numerical approach for $u_{max}=3$ and $N=9400$ discretization points, we found
about 4000 eigenenergies from $kR_0 = 75$ to $k R_0 = 150$.  By comparison with the Weyl formula (see App. \ref{sec:Weyl}), we then obtained 98.5\%  of the eigenvalues in this range. The Nearest-Neighbour Spacing Distribution
(NNSD) was calculated based on the method described in App. \ref{sec:NNSD} and is plotted in Fig. \ref{fig:analytic_trumpet}(b). It exhibits a Poissonian distribution as expected for integrable systems \cite{BGS}.\\

To conclude this section, as the tractoid is rotationally symmetric, its wave equation is separable. Its radial part solves a modified Bessel equation. Its solutions are similar to
the  wave functions of the flat disk, at least near the bottom of the effective potential, at the boundary $u=0$. This corroborates and enlightens the observation of lasing WGMs in experiments.

\section{Mode computation in the dielectric tractoid cavity}
\label{sec_6}

In the previous section, we solved the 2D wave equation for the tractoid surface. In experiments, however, the microcavity has a finite thickness. Therefore this section deals with 3D passive numerical simulations which take into account the thickness and the dielectric boundary of a realistic tractoid resonator. We also consider a disk with a small thickness (ie. a flat cylinder) for comparison.

Numerical simulations of the passive system were performed with a homemade 3D Finite Difference
Time Domain (FDTD) code. The radius is $R_0 = 8$ µm and the thickness is $\delta = 0.15$ µm for both cavities, and the height is $h=4\,\mu$m for the tractoid. As we investigate modes with a wavenumber $\Reel(k R_0) \approx 100$, we expect a single vertical excitation in the thickness.

\begin{figure}
    \includegraphics[width = 1\linewidth]{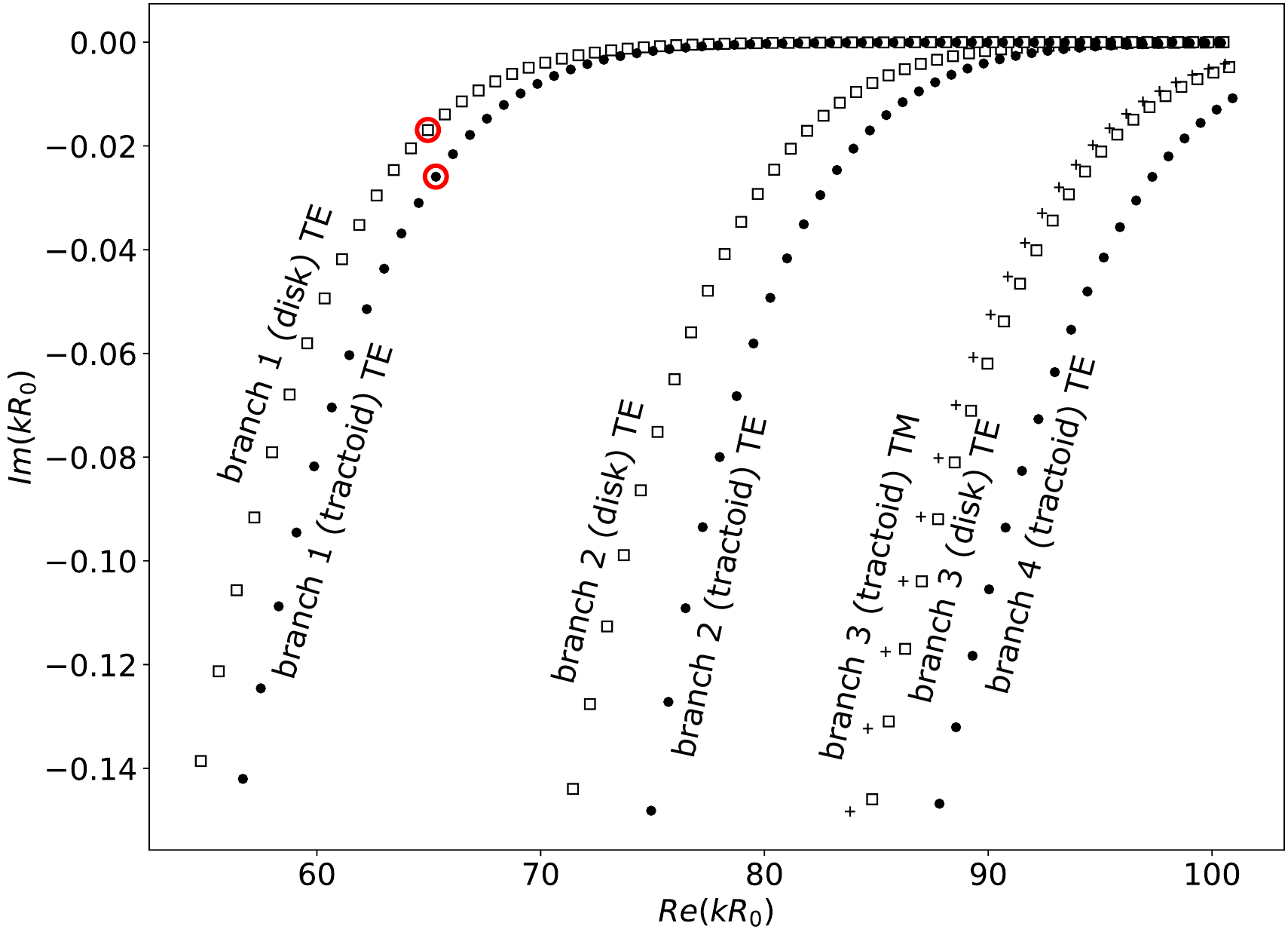}
    \caption{\label{fig:Spectre_compa} Numerical spectrum calculated by 3D FDTD
 simulations for a tractoid cavity (dots for the TE modes and crosses for the
 TM modes) and a disk cavity (empty squares). The modes ($m=70$, $p=1$) for the disk and the tractoid are circled in red.}
\end{figure}

\begin{figure*}
    \includegraphics[width=16cm]{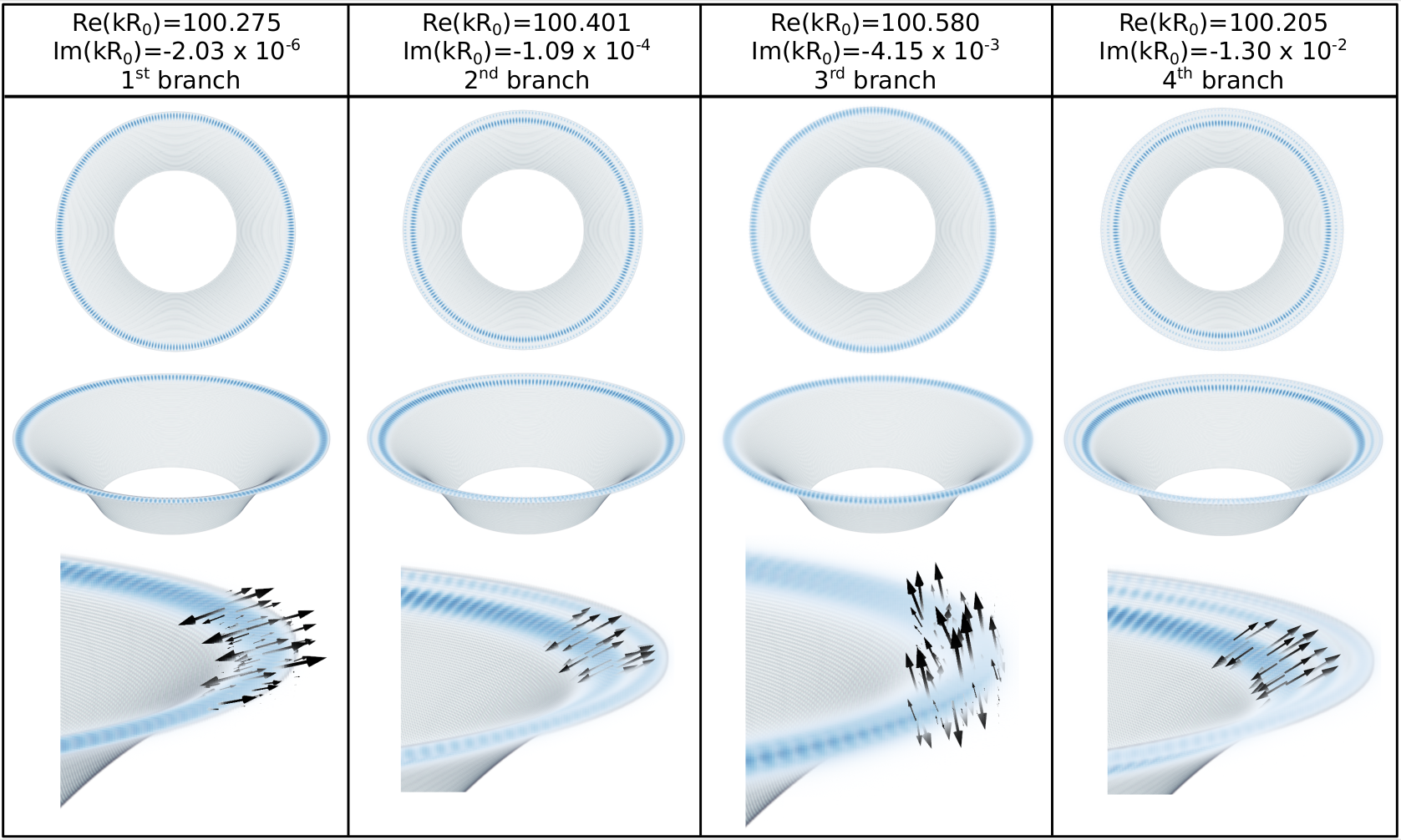}
    \caption{\label{fig:modes_FDTD} Top and side views of $|\vec{E}|^{2}$ for four modes calculated for the tractoid by 3D FDTD simulations. A source at the central frequency Re$(kR_0)=100.392$ is used for the first three columns, and at Re$(kR_0)=100.205$ for the fourth column, with a frequency width Re$(kR_0)=0.3351$. Each mode belongs to a different branch of the
 spectrum of Fig. \ref{fig:Spectre_compa}. The bottom row shows a
 magnifications near the boundary of the tractoid where the electric field polarization is
 indicated by black arrows.}
\end{figure*}

Due to their rotational symmetry, the modes of the disk and the tractoid are two-fold degenerate. Therefore we restrict the simulation domain to half a tractoid (and half a disk) and to even wave functions, which corresponds to  imposing Neumann boundary conditions along the plane $X=0$. The source is a linear dipole positioned in the plane $X=0$ at $u=0.23$, which is very close to the boundary. In Figs. \ref{fig:Spectre_compa} and \ref{fig:modes_FDTD}, the source is oriented parallel to the $Y$ axis, and thus excites preferentially TE-polarized modes. However if the source is parallel to the $Z$ axis, the simulation does not yield any useable signal.

The computed spectra of the tractoid and the disk
are plotted in Fig. \ref{fig:Spectre_compa}. The resonance wavenumbers are organized in several branches. To get further insight into the mode structure, we compute the mode profile of a given resonance using a source with a very narrow frequency range around the central frequency of this specific mode. It is inferred from the mode profiles that each branch of the tractoid spectrum corresponds to a given radial excitation $p$, with increasing $m$ azimuthal numbers, in full similarity with the disk spectrum. For a given pair ($m$, $p$), the resonance wavenumber $k_{m,p}$ of the tractoid is slightly shifted towards higher frequencies and higher losses compared to the disk. For illustration, the modes ($m=70$, $p=1$) of the disk and the tractoid are circled in red in Fig. \ref{fig:Spectre_compa}.

Fig. \ref{fig:modes_FDTD} presents the four modes inferred from the tractoid simulations with a source at a
central frequency Re$(kR_0)\simeq100$. The profiles of these wave functions are in complete agreement with
the results of Sec. \ref{sec_5}. Each mode belongs to a different branch of the spectrum of Fig.
\ref{fig:Spectre_compa}. The modes of branches 1, 2, and 4 are TE-polarized with $p=1$, 2, and 3 radial
excitations respectively, while branch 3 is TM-polarized with a single radial excitation $p=1$.
This splitting of polarizations is expected since the tractoid surface fulfills the corresponding
assumptions of the model in Ref. \cite{peschel}. For the disk, only TE-modes are excited with such a source.

In experiments, TM-modes are not lasing, which can be explained by the higher losses of
 TM-modes compared to TE-modes, as shown in Fig. \ref{fig:Spectre_compa}.\\

\begin{figure}
    \includegraphics[width = 1\linewidth]{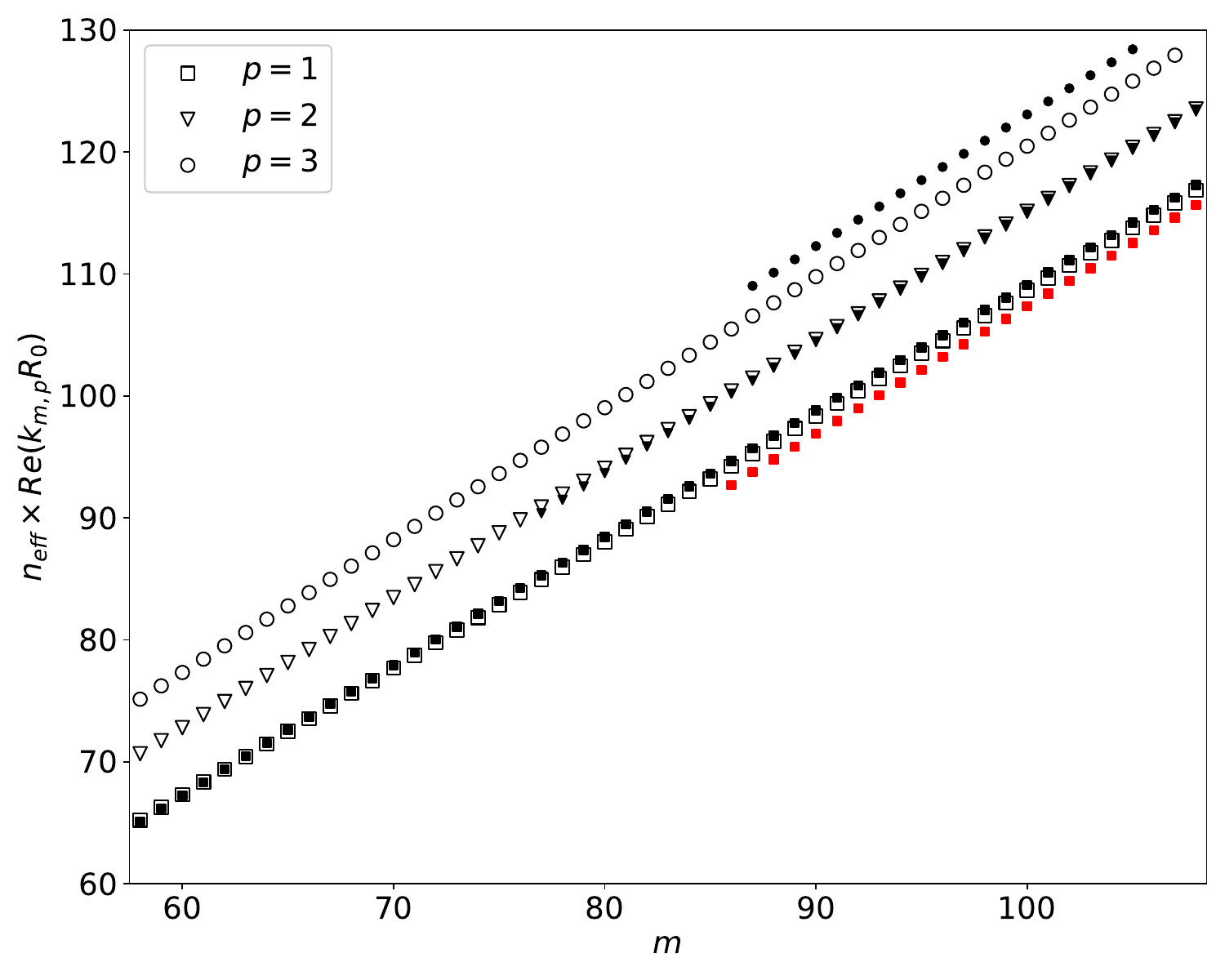}
    \caption{\label{fig:compa-theory-FDTD} Comparison of the quantity $n_{\textrm{eff}}$Re($kR_0$) between
    the analytic spectrum obtained in Sec. \ref{sec:num-dirichlet} using $n_{\textrm{eff}}=1$ (empty markers)
    and the numerical spectrum computed in Sec. \ref{sec_6} by 3D FDTD using the usual effective index
    formula, for instance Eq. (4) of \cite{PRA2007} (filled markers, TE modes in black and TM modes in red).}
\end{figure}

Finally, Fig. \ref{fig:compa-theory-FDTD} shows that this 3D numerical spectrum is very close to the analytic
spectrum obtained in Sec. \ref{sec:analytique-comparaison} and \ref{sec:num-dirichlet}, which validates
\emph{a posteriori} the use of Dirichlet boundary conditions in the analytic treatment, consistent high
quality factor WGMs \cite{commentaire-p=3}.

\section{Conclusion}
\label{sec_7}
Organic tractoid microlasers were investigated both experimentally and theoretically. We
evidenced that, thanks to its rotational symmetry, the tractoid shares many ray
and wave properties with the 2D flat disk, despite its constant negative curvature.

From the level of ray dynamics, we calculated the monodromy matrices for propagation and reflection on a constant negative surface, and then showed that all periodic geodesics on the tractoid are marginally stable. From a wave point of view,
we have numerical and experimental evidences that lasing modes are indeed WGMs. We noticed that modes exhibit two polarization states, namely TE- and TM-polarization, similar to flat cavities.
Finally,  the scalar wave equation
 can also be solved analytically leveraging on the rotational symmetry to separate variables.

Working with the Poincaré
half-plane representation significantly simplifies the computations, but most
methods discussed in this paper can also be applied to any surface of revolution.

In summary, the tractoid is integrable  because of its rotational symmetry and in spite of the exponential divergence of ray trajectories induced by the negative curvature. When breaking the rotational symmetry, however, chaotic dynamics is expected \cite{bogomolny}, which will be the focus of a further exploration in non-Euclidean photonics.

\acknowledgments

This work was done with the C2N micro nanotechnologies platform. It was supported by the French RENATECH network, by the departement de l'Essonne, and by the French state funds managed by the ANR within the Investissements d'Avenir programme under reference ANR-19-CE09-0017 FRONTAL. S.B. acknowledges support for the Chair in Photonics from R\'egion Grand-Est, D\'epartement de la Moselle, European Regional Development Fund,  CentraleSup\'elec, Fondation CentraleSup\'elec, and the Eurometropole de Metz. B.D. acknowledges financial support from the Institute for Basic Science in Korea through the project IBS-R024-D1.\\

J.-F. Audibert is acknowledged for the fabrication of centimeter-scale tractoids with a 3D printer, Y.-H. Melka and O. Ivashtenko for preliminary works on spherical squares, A. Comtet for Sec. \ref{sec:analytique-comparaison}, and S. Nonnenmacher for Fig. \ref{fig:compa-theory-FDTD}.

\appendix

\section{Bulk, effective, and group indices}\label{sec:indice}

This appendix discusses the different refractive indices relevant for the microlasers presented here.
\begin{itemize}
\item The \emph{bulk index} is the refractive index of the material itself. The bulk index of the resist IP-G was measured 1.51 at a wavelength of 600 nm in Ref. \cite{indice}. It is probably slightly different in our experiments due to the addition of the laser dyes and to the fabrication process. But as this index is difficult to measure precisely and its variation does not change significantly the results, we used the value 1.51 for the 3D simulations of Sec. \ref{sec_6}.
\item The \emph{effective index} is the refractive index which is experienced by the light propagating in a guiding layer. This is why it is involved in Eq. (\ref{eq:vec_helm2D}). The effective index is lower than the bulk index, since it considers the part of the wave which propagates outside the guiding layer (that is in air with $n=1$ in our case). The effective index can be calculated depending on the geometry of the device, on the bulk indexes of the involved layers, and on the polarization. In App. V.A of Ref. \cite{Moebius}, we showed that the curvature of the propagating layer does not modify the effective index for the range of parameters of our experiments. For instance, the effective index is 1.5 at 600 nm for TE modes and a thickness of 1 $\mu$m for the most confined modes.
\item The \emph{group index} takes into account the dispersion and must then be involved in formula (\ref{eq:fsr}) which considers the spacing between different wavelengths. A demonstration is given in \cite{3DNina} for instance. A calibration protocol is detailed in App. V.B of Ref. \cite{Moebius} and yields $n_g=1.58\pm0.05$ for our experiments.
\end{itemize}

\section{Fourier transform versus Lomb-Scargle periodogram}\label{sec:Lomb}

In this Appendix we discuss the sampling issue in Fourier transform calculations.

The Discrete Fourier Transform (DFT) from discrete times $t_n$ to frequency $f$ is defined as
\begin{equation}\label{eq:DFT-basique}
F(f)=\sum_{n=1}^{n=N}y_n\,e^{-2i\pi ft_n}
\end{equation}
where the data $y$ is sampled at the times $t_n$, yielding a vector $(y_n)$ of size $N$. Most DFT algorithms assume that the sampling of the data is regular, which means that $t_{n+1}-t_{n}$ is a constant. However our experimental spectrum is not regularly sampled for the following two reasons:
\begin{itemize}
\item The pixels of the spectrometer camera are regularly spaced, but not the corresponding wavelengths $\lambda_n$ because the grating of the spectrometer induces a non-linear dispersion.
\item The Fourier transform has to be calculated as function of the wavenumbers $k_n$ instead of the wavelengths $\lambda_n=2\pi/k_n$ and this relation is not linear either.
\end{itemize}
Therefore, $\Delta_n= k_{n+1}-k_n$ is not constant for the experimental spectra. In the top panel of Fig. \ref{fig:tractoid_exp_result}(b), we plot the Fourier transform of the tractoid experimental spectrum of Fig. \ref{fig:tractoid_exp_result}(a) calculated by the FFT algorithm implemented on Python in the \texttt{scipy} library, which corresponds to the modulus squared of formula (\ref{eq:DFT-basique})  assuming a constant sampling $\Delta=(k_{2}-k_1)/N$. For comparison, in the middle panel of Fig. \ref{fig:tractoid_exp_result}(b), we plot the modulus square of expression (\ref{eq:DFT-basique}) using the actual $(k_n)$, which are not equidistant. Note that the main peak is significantly shifted, revealing an artefact: the raw DFT (top panel) is peaked at a length which is longer than the perimeter and thus not consistent with WGMs, while in the middle panel the length is slightly shorter than the perimeter, in agreement with WGMs.\\

The Lomb periodogram $P_{LS}(f)$ plotted in the bottom panel of Fig. \ref{fig:tractoid_exp_result}(b) was computed by the \texttt{lombscargle} routine implemented on Python in the \texttt{scipy} library. Its expression is given in \cite{lomb-python} and is similar to the formula in the seminal paper by \cite{Lomb}. It is peaked at the same length as the well-sampled DFT in the middle panel of Fig. \ref{fig:tractoid_exp_result}(b),  however, the signal-to-noise ratio is much better than for the other two procedures. Therefore, we prefer the Lomb periodogram for analyzing the experimental spectra.


\section{Stability computations of periodic geodesics}
\label{appendix:stability}

This Appendix deals with the stability of geodesics on the tractoid surface. First we derive the monodromy matrix for propagation $\mathcal{P}$, then the monodromy matrix for reflection $\mathcal{R}$. Finally in App. \ref{subappendix:tractoide-complete} and \ref{subappendix:tractoide-tronquee} we apply these formulas to periodic geodesics of the full and truncated tractoid, respectively.

\subsection{Monodromy matrix for propagation}
\label{sec:stabilite-matrices}
\begin{figure}[tb]
    \includegraphics[width=6cm]{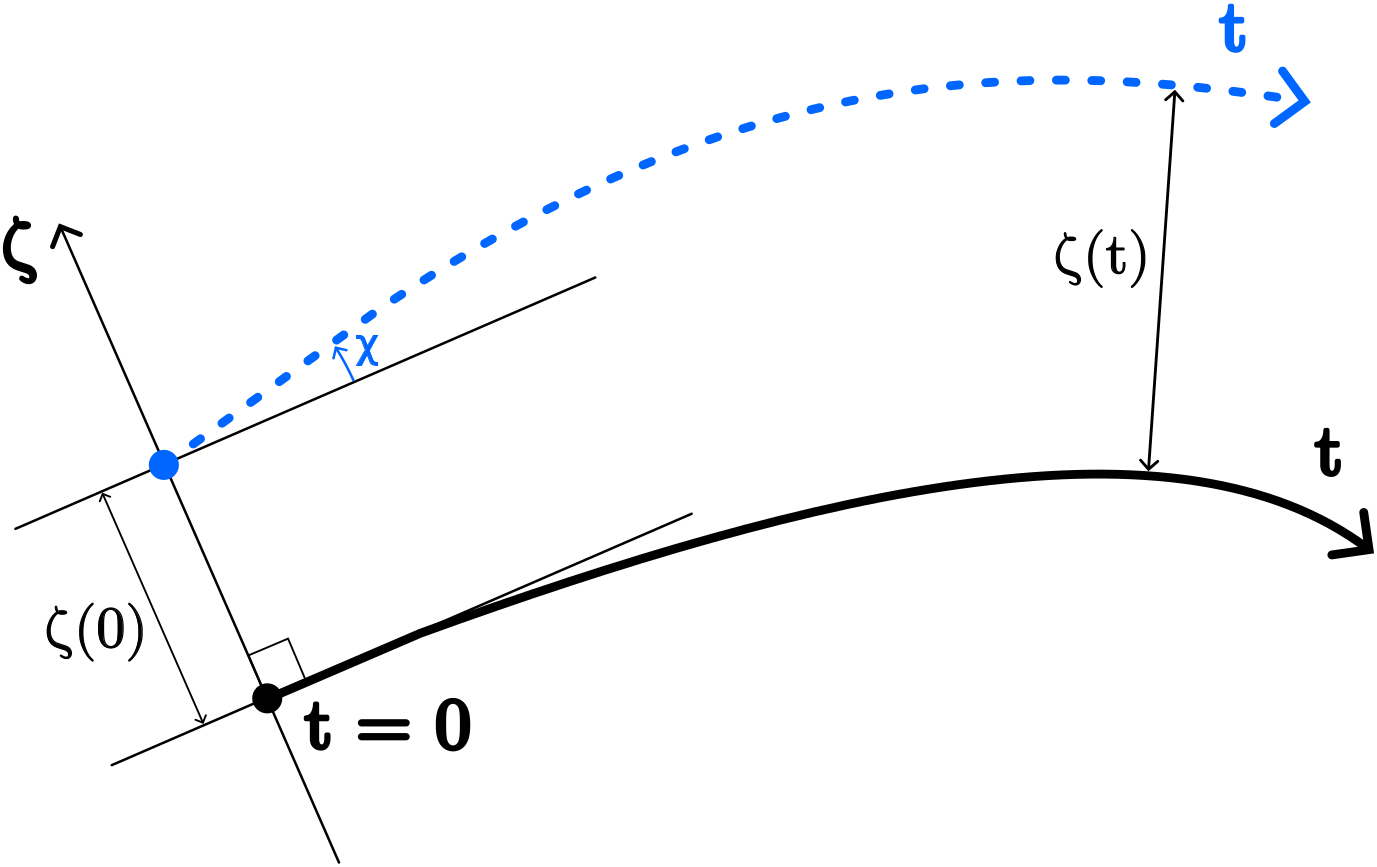}%
    \caption{\label{fig:monodromy_schema} Schematic for the computation of the
 propagation monodromy matrix. The black curve is the reference geodesic. It is
 parametrized by $t$ such that the black dot corresponds to $t=0$. The blue dotted line
 is the perturbated geodesic. The initial point at $t=0$ is represented
 by the blue dot and is shifted by a distance $\zeta(0)$. The initial angle of the
 blue curved with respect to the reference is called $\chi$, which is equal to $\zeta'(0)$.}
\end{figure}

In this paragraph, we derive the monodromy
matrix $\mathcal{P}(t)$ for the propagation along a geodesic on the tractoid surface. The usual
$(\zeta, \sin(\chi))$ coordinate system along the unperturbed trajectory is convenient, where $\zeta$ is the coordinate in a direction orthogonal to the geodesic tangent vector and $\chi$ is the deviation from the original propagation direction, as shown in Fig. \ref{fig:monodromy_schema}.
The coordinate $\sin(\chi)$ corresponds
to $d\zeta / dt$, ie. the local slope with respect to the unperturbed geodesic. Within this linearized stability analysis,
the matrix $\mathcal{P}(t)$ is defined as
\begin{equation}\label{eq:def-pi}
    \begin{pmatrix}
        \zeta(t) \\ \zeta'(t)
    \end{pmatrix}
	= \mathcal{P}(t) \begin{pmatrix}
        \zeta(0) \\ \zeta'(0)
    \end{pmatrix}
\end{equation}
The Jabobi equation \cite{kuehnel} describes the evolution of $\zeta$ along a given geodesic as a function of $t$ and reads
\begin{equation}
 \frac{d^{2}\zeta}{dt^{2}} + K \,\zeta = 0
\end{equation}
where $K$ is the Gaussian curvature. As it is a constant equal to
$-1/R_0^{2}$ for the tractoid, the general solution follows
\begin{equation}\label{eq:zeta}
    \zeta(t) = \zeta(0) \cosh\left(\frac{t}{R_0}\right) +
     \zeta'(0)R_0 \sinh\left(\frac{t}{R_0}\right)
\end{equation}
Calculating the derivative of Eq.~(\ref{eq:zeta}) and comparing it with (\ref{eq:def-pi}) yields
\begin{equation}
	\mathcal{P} (t) =
    \begin{pmatrix}
 \cosh \left( \frac{t}{R_0} \right) & R_0 \sinh \left( \frac{t}{R_0} \right) \\
 \frac{1}{R_0} \sinh \left( \frac{t}{R_0} \right) & \cosh \left( \frac{t}{R_0} \right)
    \end{pmatrix}
\end{equation}
which is the monodromy matrix for propagation along a geodesic on the tractoid. As expected, it has the property $\mathcal{P}(t_1)\mathcal{P}(t_2)=\mathcal{P}(t_1+t_2)$ and it tends to the propagation matrix in a planar billiard in the limit of vanishing Gaussian curvature $R_0\to\infty$.
Note that $\det (\mathcal{P})=1$ and $\tr(\mathcal{P})=2 \cosh(\frac{t}{R_0})$
which is larger than 2. Therefore, one of its eigenvalues is larger than
1 and the propagation is unstable as it is expected on a surface with a negative curvature.

\subsection{Monodromy matrix for reflection}
\label{subappendix:stability2}
In this section, we derive the monodromy matrix for the reflection of a ray on the surface boundary,
following Ref. \cite{sieber2}. First, the general case of a given surface and a given
boundary is studied. Then this result is applied to the specific case of the tractoid.

\subsubsection{The general case}
\label{subappendix:stability2_1}
The surface is defined by the vector $\vec{F}(u, v)$ [cf. Eqs. (\ref{eq:Fx}-\ref{eq:Fz})] and parameterized by the
variables $u$ and $v$. The boundary is a curve belonging to the surface and defined
by the vector $\vec{c}\,(t)$. The parametrization is chosen such that
$\| \frac{d \vec{c}}{dt}\| = 1$. This choice simplifies the definition of the
curvature of the curve which is then the modulus  of the second derivative:
$\kappa = \| \frac{d^{2} \vec{c}}{dt^{2}}\|$.

\medskip
In the vicinity of the reflection point, the incident and reflected geodesics belong
to the plane which is tangent to the surface. This brings us back to the case of the reflection in the Euclidean plane. The monodromy matrix for the reflection with an incident angle $\chi$ with the surface normal is
therefore
\begin{equation}
    \mathcal{R}(\chi) =
    \begin{pmatrix}
        -1 & 0 \\
        \frac{2\epsilon \kappa_t}{\cos \chi} & -1
    \end{pmatrix}
\end{equation}
where $\epsilon=\pm 1$ depends on the relative position of the light ray and the boundary, and $\kappa_t$ is the tangent curvature (also known as the geodesic curvature).
The parameter $\epsilon$ can be defined as
\begin{equation}
\epsilon =\textrm{sign}\left(\vec p\,.\,\frac{d^{2} \vec{c}}{dt^{2}}\right)
\end{equation}
where $\vec p$ is the momentum of the ray \emph{after} reflection on the boundary.

It remains to calculate  the tangent curvature $\kappa_t$. To formalize that, we denote $\vec{N}$ the unit
vector normal to the surface. If the curvature of the boundary $\frac{d^{2} \vec{c}}{dt^{2}}$ is collinear
to $\vec{N}$ then all the curvature is outside the tangent plane and thus the tangent curvature $\kappa_t$
is zero. This is the case when the curve $\vec{c}$ is a geodesic. Hence, a reflection on a geodesic is
equivalent to a reflection on a plane mirror in the 2D Euclidean case.

In general, the vector normal to the surface $\vec{N}$ and the curvature of the boundary $\vec{c}$ are not
collinear. The tangent curvature $\kappa_t$ is the modulus of the projection of
$\frac{d^{2} \vec{c}}{dt^{2}}$ onto the tangent plane:
\begin{equation}
    \kappa_t = \left \Vert \frac{d^{2} \vec{c}}{dt^{2}} - \left(\frac{d^{2} \vec{c}}{dt^{2}}
    .\vec{N}\right) \vec{N}\right \Vert
\end{equation}
which is equivalent to
\begin{equation}
    \kappa_t = \left \Vert \frac{d^{2} \vec{c}}{dt^{2}} \cdot \left(
    \frac{d \vec{c}}{dt} \times \vec{N} \right)\right \Vert
    \label{eq:kappa_general}
\end{equation}
since $(\frac{d\vec c}{dt}, \vec N, \frac{d\vec c}{dt}\times \vec N)$ is a direct orthonormal basis, and $\frac{d\vec c}{dt}$ is orthogonal to $\frac{d^{2} \vec{c}}{dt^{2}}$.

\subsubsection{The case of the tractoid}
\label{subappendix:reflexion-tractoide}
In this paragraph, we consider the specific case where the tractoid is cut at a constant coordinate $u_0$. Then the boundary is a horizontal circle of radius $R = \frac{R_0}{\cosh(u_0)}$. Choosing the
parametrization to satisfy $\| \frac{d \vec{c}}{dt}\| = 1$, we obtain
\begin{align}
 \vec{c}(t) & = \left( R \cos \left( \frac{t}{R}\right), R \sin \left( \frac{t}{R}\right), R_0(u_0-\tanh u_0) \right) \\
 \frac{d^{2} \vec{c}}{dt^{2}} & = -\frac{1}{R} \left( \cos \left( \frac{t}{R}\right), \sin \left( \frac{t}{R}\right), 0 \right)
\end{align}
with $t\in[0,2\pi[$.
Moreover, the unit vector normal to a surface is given by
\begin{equation}
 \vec{N} = \frac{\frac{\partial \vec{F}}{\partial u} \times \frac{\partial \vec{F}}
 {\partial v}}{\left|\left| \frac{\partial \vec{F}}{\partial u} \times \frac{\partial \vec{F}}
 {\partial v} \right|\right|}
\end{equation}
where $\vec{F}(u,v) = (F_X(u,v), F_Y(u,v), F_Z(u,v))$ is the surface equation
given in formulas (\ref{eq:Fx}-\ref{eq:Fz}). For the tractoid,
it yields
\begin{equation}
 \vec{N} = \frac{1}{\cosh(u_0)}
    \begin{pmatrix}
 \sinh(u_0) \cos(v) \\ \sinh(u_0) \sin(v) \\ 1
    \end{pmatrix}
\end{equation}
Due to rotational invariance, the tangent curvature $\kappa_t$ does not depend on the coordinate $v$, so we can take $v=0$ and after computation using formula (\ref{eq:kappa_general}),
we find $\kappa_t=\frac{1}{R_0}$ independent of $u_0$.
Therefore, for any $u_0$, the monodromy matrix for the reflection is identical to the reflection matrix on a 2D circular boundary of
radius $R_0$.

\begin{equation}
    \mathcal{R}(\chi) =
    \begin{pmatrix}
        -1 & 0 \\
        \frac{2\epsilon}{R_0 \cos \chi} & -1
    \end{pmatrix}
\end{equation}
with $\epsilon = +1$ for the reflection on the tractoid boundary $u=0$, while it is $\epsilon =-1$ for $u=u_{max}$.

\subsection{Stability computations for the full tractoid}
\label{subappendix:tractoide-complete}
In this paragraph, the full infinite tractoid is considered  for which only the
periodic geodesics described in Fig. \ref{fig:schema_PG} exist. Using the same notations as in Sec.
\ref{sec_4}, the total monodromy matrix for the propagation of the $(M, P)$-periodic geodesic can be written
\begin{equation}
    \mathcal{M}(M, P) = [\mathcal{M}_0 (M, P)]^{M}
\end{equation}
with $\mathcal{M}_0 (M, P) = \mathcal{P}(M, P) \mathcal{R}(M, P)$ the product of the propagation matrix and the
reflection matrix for the $(M,P)$-periodic geodesic. Combining formulas
(\ref{eq:length_PG}) and (\ref{eq:angle_PG}) with the expression of the matrices in Eqs.
 (\ref{eq:matrice-pi}) and (\ref{eq:matrice-R}),
we obtain the following expressions for $\mathcal{P}(M, P)$

\begin{eqnarray*}
	\mathcal{P} (M, P) = \begin{pmatrix}
        \Pi_{1} & R_0 \Pi_{2} \\
        \frac{\Pi_{2}}{R_0} & \Pi_{1}
    \end{pmatrix}
\end{eqnarray*}
with
\begin{align*}
    \Pi_{1} & = 2\left(\frac{P \pi}{M}\right)^{2} + 1 \\
    \Pi_{2} & = 2 \left(\frac{P \pi}{M}\right)
 \left( 1 + \left(\frac{P \pi}{M}\right)^{2} \right)^{\frac{1}{2}}
\end{align*}
and for $\mathcal{R}(M, P)$
\begin{equation}
 \mathcal{R}(M, P) =
    \begin{pmatrix}
 -1 & 0 \\
 \frac{2}{R_0} \frac{M}{P \pi} \left( 1 + \left(\frac{P \pi}{M}\right)^{2} \right)^{\frac{1}{2}} & -1
    \end{pmatrix}
\end{equation}
The computation of the trace of $M_0(M, P)$ is then cumbersome but straightforward
and yields $\tr (M_0)= 2$. Thus, these periodic geodesics are
marginally stable.

\subsection{Stability computations for the truncated tractoid}
\label{subappendix:tractoide-tronquee}
In this subsection, we consider a truncated tractoid. Once again, it is much more convenient to work in the Poincaré half-plane because the lower boundary at $u=0$ and the
top boundary at $u = u_{max}$ are respectively
mapped to the straight horizontal lines $y=1$ and $y=y_{max}= \cosh(u_{max})$. The periodic
geodesics studied in Sec. \ref{subsec_4_2} still exist provided that $u_{max}$ is high enough, ie. that $\theta_{M, P}$ is smaller than a maximal angle $\theta_{max} = \arccos\left(
\frac{1}{y_{max}}\right)$. They remain marginally stable.

\begin{figure}
    \includegraphics[width=8.5cm]{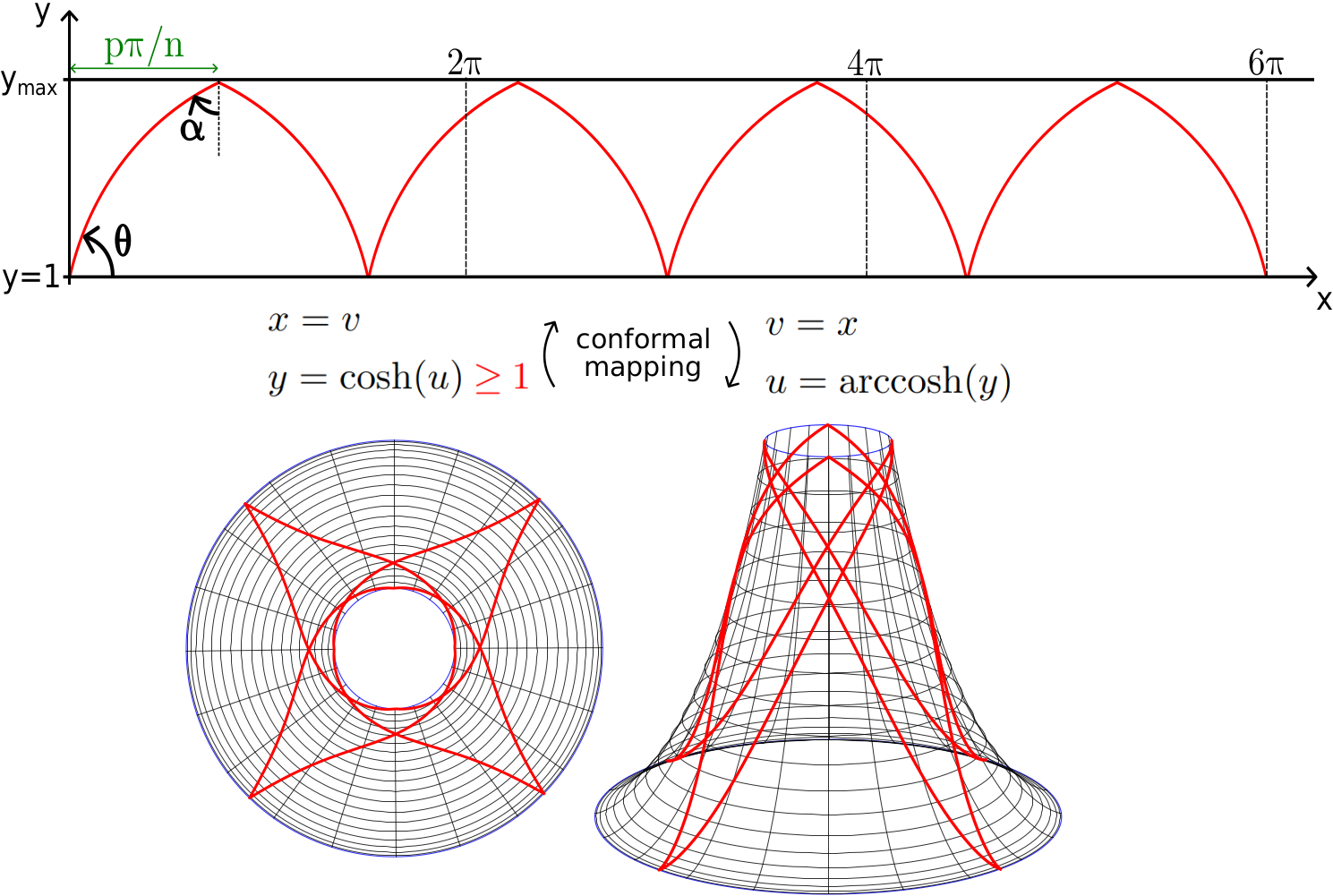}%
    \caption{\label{fig:schema_PG_type2} Periodic geodesic for the truncated tractoid with $(M, P) = (4, 3)$ in the Poincaré half-plane (top) and on the actual tractoid (bottom).}
\end{figure}

However, rays can also be reflected on the
top boundary and therefore a new type of periodic geodesic appears. Since the
top boundary is also a straight horizontal line, the reflection angles are
 preserved. It entails that such periodic geodesics are composed of
a series of identical patterns, made of circular arcs as illustrated in Fig. \ref{fig:schema_PG_type2}. Each such periodic geodesic is uniquely identified by a pair of positive relatively
prime integers that will also be called $(M, P)$, $M$ being the number of reflections on the
lower boundary and $P$ the rotation number. The $(4,3)$ periodic geodesic is sketched in
Fig. \ref{fig:schema_PG_type2}.

Once again, $L_{M, P}$ is the total length of the $(M, P)$ periodic
geodesics, $\theta_{M, P}$ the complementary angle of the incident angle on
the lower boundary and $\alpha_{M, P}$ the incident angle on the upper boundary
(cf. Fig. \ref{fig:schema_PG_type2} for the notations). After simplification,
we obtain
\begin{align}
    L_{M, P} & = 4 M R_0 \arcsinh \left( \frac{1}{2} \sqrt{\frac{\left( \frac{P}{M} \pi \right)^{2}
    + (y_{max}-1)^{2}}{y_{max}}} \right) \\
    \theta_{M, P} & = \arctan \left( \frac{y_{max}^{2}-1}{2 \frac{P}{M} \pi}
    + \frac{P}{M} \frac{\pi}{2} \right) \\
    \alpha_{M, P} & = - \arctan \left( \frac{y_{max}}{\frac{y_{max}^{2}-1}{2 \frac{P}{M} \pi}
    - \frac{P}{M} \frac{\pi}{2}} \right)
\end{align}
As previously, the monodromy matrix of the $(M, P)$ periodic
geodesics $\mathcal{M}'(M, P)$ can be written
\begin{equation}
    \mathcal{M}'(M, P) = [\mathcal{M}'_0 (M, P)]^{M}
\end{equation}
The matrix $\mathcal{M}'_0 (M, P)$ is a product of
matrices for propagation and reflection of the form
\begin{equation}
	\mathcal{M}'_0 (M, P) = \mathcal{P}(M, P) \mathcal{R}(M, P) \mathcal{P}(M, P) \tilde{\mathcal{R}}(M, P)
\end{equation}
where $\mathcal{P}(M, P)$ accounts for the propagation from the upper boundary to the lower one
or the opposite, $\mathcal{R}(M, P)$ for the reflection at the boundary $y=1$ and
$\tilde{\mathcal{R}}(M, P)$ for the reflection at the boundary $y=y_{max}$. Note that $\epsilon =+1$ in $\mathcal{R}(M, P)$, and $\epsilon =-1$ in $\tilde{\mathcal{R}}(M, P)$.\\
After cumbersome but trivial calculations,  we obtain that the trace of $M'_0(M, P)$ is equal to 2. These periodic geodesics are thus also marginally stable as expected. It should be noted that all the periodic geodesics discussed in this article come in families since they are invariant under rotation (that is, a shift of the starting coordinate $v$).

\section{General wave equation for a surface of revolution}
\label{appendix:wave_eq_SOR}
In sec. \ref{sec_5}, the wave equation is solved using the Poincaré half-plane
representation for its simplicity. Nevertheless, the wave equation can also be derived directly
from the actual tractoid surface using separation of variables. This method applies to any Surface Of Revolution (SOR).\\

The wave equation is the scalar Helmholtz equation
\begin{equation}
 (\Delta + k^{2}) \psi = 0
\end{equation}
where $k$ is the wavenumber in vacuum and $\Delta$ is the 2D scalar Laplacian in
any coordinate system. The metric tensor for a SOR can always be written
\begin{equation}
 ds_{SOR}^{2} = E(u)du^{2}+G(u)dv^{2}
\end{equation}
Therefore, the 2D scalar Laplacian $\Delta$ in the $(u, v)$ local coordinate
system is
\begin{equation}
    \Delta \psi = \frac{1}{\sqrt{E(u)G(u)}} \frac{\partial}{\partial u} \left(
 \sqrt{\frac{G(u)}{E(u)}} \frac{\partial \psi}{\partial u}\right) +
 \frac{1}{G(u)} \frac{\partial^{2} \psi}{\partial v^{2}}
\end{equation}
The general form of the wave equation for a SOR becomes
\begin{equation}
 \frac{1}{\sqrt{EG}} \partial_u \left( \sqrt{\frac{G}{E}} \partial_u \Phi
 \right) + \frac{1}{G} \partial_v^2 \Phi + k^2 \Phi = 0
\end{equation}
Using the separation ansatz \(\Phi(u, v) = f(u) g(v)\), we obtain
\begin{equation}
    \sqrt{\frac{G}{E}} \frac{1}{f(u)} \frac{d}{du} \left( \sqrt{\frac{G}{E}}
    f'(u)\right) + k^{2} G + \frac{g''(v)}{g(v)} = 0
\end{equation}
which shows that the variables $u$ and $v$ are indeed separable. The function $g$ must be \(2\pi\)-periodic, hence  $g(v) = C e^{imv}$ with $C \in \mathbb{C}$ and $m \in \mathbb{Z}$.
Finally, $f$ is solution of the following equation
\begin{equation}
    \sqrt{\frac{1}{EG}} \frac{d}{du} \left( \sqrt{\frac{G}{E}}
    f'(u)\right) + \left[ k^{2} - \frac{m^{2}}{G} \right] f(u) = 0
\end{equation}
For the tractoid, $E$ and $G$ are replaced by their expressions in Eqs. (\ref{eq:G}) and (\ref{eq:E}). The wave equation then becomes
\begin{equation}
 f'' - \frac{1}{\tanh u} f' + \left( \frac{(k R_0)^2}{\cosh^2 u} - m^2 \right) \sinh^2 u\, f = 0
\end{equation}
The following variable and function changes
\begin{eqnarray}
    \eta & =& \ln (\cosh u) \\
    f(\eta) & = &e^{\frac{\eta}{2}}\,\xi(\eta)
\end{eqnarray}
lead to the Schrödinger-like equation (\ref{eq:schrodinger_eq}). Therefore this
wave equation can be solved with the procedure described in Sec.
\ref{sec_5}.

\section{Matrices for solving the scalar wave equation}
\label{appendix:der_wave_eq}

In this Appendix, we provide some elements to solve the scalar wave equation of Sec. \ref{sec:num-dirichlet} via an eigenvalue problem.\\
We start by spatially discretizing the
problem. We consider $N+2$ equally spaced points
$0 = \eta_{0} < \eta_{1} < ... < \eta_{N} < \eta_{N+1} = \eta_{max}$
and we note $\xi_i = \xi(\eta_i)$ for $i \in [\![0, N+1 ]\!]$.
 For a sufficiently
small spatial step $h = \frac{\eta_{N+1} - \eta_0}{N+1}$, the second order derivative
is well approximated by the discretization
\begin{equation}
    \frac{d^{2}\xi}{d\eta^{2}} \Bigr|_{\eta=\eta_i} = -
    \frac{- \xi_{i-1} + 2 \xi_{i} - \xi_{i+1}}{h^{2}} + \mathcal{O}(h^{2})
    \label{eq:discret}
\end{equation}
for $i \in [\![1, N ]\!]$.

The following matrices are then used:
\begin{equation}
	\mathcal{T} = \frac{1}{h^{2}}\begin{pmatrix}
        2 &   -1    & 0       &   \dots      & 0              \\
       -1 &    2    & -1      &              &                \\
          &         & \ddots  &              &                \\
          &         & -1      & 2            & -1             \\
        0 & \dots   &  0      & -1           & 2
        \end{pmatrix}
\end{equation}
\begin{equation}
	\mathcal{V} = m^{2}
    \begin{pmatrix}
      e^{2 \eta_1} &              &                  \\
                      & \ddots       &                  \\
                      &              & e^{2 \eta_N}
    \end{pmatrix}
\end{equation}

\section{Weyl law}
\label{sec:Weyl}

The average number of eigenmodes $N_{smooth}(x)$ in an energy interval $[0,x]$ with $x := kR_0$ is given by the Weyl law \cite{voros},
\begin{equation}
N_{Weyl}(x)=\frac{\mathcal{A}}{4\pi}x^2-\frac{\mathcal{P}}{4\pi}x-\frac{\mathcal{A}}{12\pi}
\end{equation}
where $\mathcal{P}$ and $\mathcal{A}$ are the perimeter of the tractoid in units of $R_0$, and the area of the tractoid in units of $R_0^{2}$, respectively. For the full tractoid (i.e. $v\in[0,2\pi[$), each mode is two-fold degenerate. We consider the half-tractoid (i.e. $v\in[0,\pi[$) and obtain
\begin{eqnarray}
\mathcal{P}&=&\pi \left(1+\frac{1}{y_{max}}\right)+\ln(y_{max})\\
\mathcal{A}&=&\pi \left(1-\frac{1}{y_{max}}\right)
\end{eqnarray}
\section{Nearest-Neighbour Spacing Distribution}
\label{sec:NNSD}

The Nearest-Neighbour Spacing Distribution $P(s)$ is the probability to find a spacing of two adjacent
eigenvalues in an interval $[s,s+ds]$. For a proper comparison with the BGS conjecture \cite{BGS} which
makes statements about universal spectral properties, we need to extract the $k$-dependence of the eigenmode
mean density, which is deduced from Weyl's law (see App. \ref{sec:Weyl}). This process is referred to as
\emph{unfolding} \cite{schmit}. It leads to a mean spacing equal to 1, while the statistical properties
of the fluctuations are preserved.

The unfolded eigenvalues $\tilde{x}_i$ are obtained by replacing the eigenvalues $x_i$ by Weyl's law
\begin{equation}
\tilde{x}_i=N_{Weyl}(x_i)-\frac{1}{2}
\end{equation}
Finally, the spacings between adjacent unfolded eigenvalues is given as
\begin{equation}
s=\tilde{x}_{i+1}-\tilde{x}_i
\end{equation}

The Nearest Neighbour Spacing Distribution $P(s)$ is the histogram of the spacings $s$. In Fig. \ref{fig:analytic_trumpet} it is normalized such that the area of the histogram is one.


\begin{thebibliography}{0}

\bibitem{bottle} M. P\"ollinger, D. O’Shea, F. Warken, and A. Rauschenbeutel, \emph{Ultrahigh-Q Tunable Whispering-Gallery-Mode Microresonator}, \prl  {\bf 103,} 053901 (2009).

\bibitem{tore} T. J. Kippenberg, S. M. Spillane, and K. J. Vahala, \emph{Kerr-Nonlinearity Optical Parametric Oscillation in an ultrahigh-Q Toroid Microcavity}, \prl , {\bf 93,} 083904 (2004).

\bibitem{Poincare1895} H. Poincaré, \emph{Analysis situs}, Journal de l'Ecole Polytechnique, {\bf 1} 1 (1895).

\bibitem{voros}  N. L. Balazs and A. Voros, \emph{Chaos on the pseudosphere}, Physics Reports {\bf 143,} 109 (1986).

\bibitem{sieber} R. Aurich, M. Sieber, and F. Steiner, \emph{Quantum chaos of the Hadamard-Gutzwiller model} \prl {\bf 61,} 483 (1988).

\bibitem{szepfalusy} A. Csord\'as, R. Graham, and P. Sz\'epfalusy, \emph{Level statistics of a noncompact cosmological billiard}, \pra , {\bf 44,} 1491 (1991).

\bibitem{bogomolny} E. Bogomolny, B. Georgeot, M.-J. Giannoni, and C. Schmit, \emph{Arithmetical chaos}, Physics Reports {\bf 291}, 219 (1997).

\bibitem{quantumhall} M. F\"urst, D. Kochan, I.-G. Dusa, C. Gorini, and K. Richter, \emph{Dirac Landau levels for surfaces with constant negative curvature}, \prb  {\bf 109,} 195433 (2024).

\bibitem{escher} M. C. Escher \emph{Circle limit I-IV}.

\bibitem{peschel-PRL} V. H. Schultheiss, S. Batz, A. Szameit, F. Dreisow, S. Nolte, A. T\"unnermann, S. Longhi, and U. Peschel, \emph{Optics in curved space}, \prl , {\bf 105,} 143901 (2010).

\bibitem{arie-polariton} A. Libster-Hershko, R. Shiloh, and A. Arie, \emph{Surface plasmon polaritons on curved surfaces}, Optica, {\bf 6,} 115 (2019).

\bibitem{segev} R. Bekenstein, Y. Kabessa, Y. Sharabi, O. Tal, N. Engheta,
G. Eisenstein, A. J. Agranat, and M. Segev, \emph{Control of light by curved space in nanophotonic structures}, Nat. Phot. {\bf 11, } 664 (2017).

\bibitem{Moebius} Y. Song, Y. Monceaux, S. Bittner, K. Chao, H. M. Reynoso de la Cruz,
C. Lafargue, D. Decanini, B. Dietz, J. Zyss, A. Grigis, X. Checoury, and M. Lebental, \emph{M{\"{o}}bius Strip Microlasers: A Testbed for Non-Euclidean Photonics}, \prl, {\bf 127}, 203901 (2021).

\bibitem{gutkin} B. Gutkin, U. Smilansky, and E. Gutkin, \emph{Hyperbolic Billiards on Surfaces of Constant Curvature}, Commun. Math. Phys. {\bf 208,} 65 (1999).

\bibitem{instabilite} A geodesic on a negatively curved surface is unstable regarding the propagation on the surface without boundary. Eventually, it can be stabilized through reflections on the boundary, which is here the case for the tractoid.

\bibitem{kuehnel} W. K\"uhnel, \emph{Differential geometry. Curves - surfaces - manifolds}, third edition, AMS (2015).

\bibitem{transmission} The main difference between Newton's laws and geometrical optics is the possibility for a ray of light to be partially refracted at a dielectric boundary following Fresnel's equations, and then partially transmitted again inside the cavity. But it requires a concave boundary, which is not the case for the tractoid.

\bibitem{PNASChenni} C. Xu, I. Dana, L.-G. Wang, and P. Sebbah, \emph{Light chaotic dynamics in the transformation from curved to flat surfaces}, Proc. Nat. Acad. Sci. , {\bf 119}, e2112052119 (2022).

\bibitem{berry} M. Berry and M. Tabor, \emph{Closed orbits and the regular bound spectrum}, Proc. R. Soc. Lond. A. {\bf 349,} 101 (1976).


\bibitem{brack}
M. Brack and R. K. Bhaduri, \emph{Semiclassical physics},
Westview Press, Oxford (2003).

\bibitem{sieber2} M. Sieber, \emph{Billiard systems in three dimensions: the boundary
integral equation and the trace formula}, Nonlinearity, {\bf 11,} 1607 (1998).

\bibitem{indice} T. Gissibl, S. Wagner, J. Sykora, M. Schmid, and H. Giessen, \emph{Refractive index measurements of photoresists for three-dimensional direct laser writing} Optical Materials Express, {\bf 7,} 2293 (2017).

\bibitem{PML} In the numerical simulations of Sec. \ref{sec_6} we also used Perfectly Matched Layers (PML) at a certain distance from the tractoid to absorb the outcoming waves.

\bibitem{peschel} S. Batz and U. Peschel, \emph{Linear and nonlinear optics in curved space}, \pra {\bf 78,} 043821 (2008).

\bibitem{bessel} R. B. Paris, \emph{On the $\nu$-zeros of the modified Bessel function $K_{i\nu}(x)$ of
positive argument}, arXiv:2108.01447 (2021).

\bibitem{BGS} O. Bohigas, M. J. Giannoni, and C. Schmit, \emph{Characterization of chaotic quantum spectra and universality of level fluctuation laws}, \prl {\bf 52,} 1 (1984).

\bibitem{commentaire-p=3} The agreement for $p=3$ can probably be improved taking into account the actual thickness of the tractoid cavity at the $u$ position where the wave function is actually localized, and not using $h=0.15~\mu$m which is the thickness at the edge.

\bibitem{PRA2007} M. Lebental, N. Djellali, C. Arnaud, J.-S. Lauret, J. Zyss, R. Dubertrand, C. Schmit, and E. Bogomolny, \emph{Inferring periodic orbits from spectra of simple shaped micro-lasers}, \pra {\bf 76,} 023830 (2007).

\bibitem{3DNina} N. Sobeshchuk, M. A. Guidry, C. Lafargue, R. Gashemi, D. Decanini, J. Zyss, and M. Lebental, \emph{Out-of-plane modes in three-dimensional Fabry-Perot microlasers}, \apl, {\bf 112,} 261102 (2018).

\bibitem{lomb-python} $https://github.com/scipy/scipy/blob/v1.14.1/scipy/$\\
    $signal/\_spectral.py$

\bibitem{Lomb} N. R. Lomb, \emph{Least-squares frequency analysis of unequally spaced data}, Astrophysics and Space Science, {\bf 39} 447 (1976).

\bibitem{schmit} C. Schmit, \emph{Quantum and classical properties of some billiards on the hyperbolic plane} in Les Houches, Session LII, 1989, \emph{Chaos and quantum physics}, M.-J. Giannoni, A. Voros, and J. Zinn-Justin, eds.



\end{thebibliography}
\end{document}